# Acidity-Mediated Metal Oxide Heterointerfaces: Roles of Substrates and Surface Modification

Gyu Rac Lee[1+], Thomas Defferriere[1+], Jinwook Kim[3], Han Gil Seo[4], Yeon Sik Jung[2], and Harry L. Tuller[1*]

*[1]Department of Materials Science and Engineering, Massachusetts Institute of Science and Technology, Cambridge, MA 02139, USA*
*[2]Department of Materials Science and Engineering, Korea Advanced Institute of Science and Technology, 291 Daehak-ro, Yuseong-gu, Daejeon 34141, Republic of Korea*
*[3]Department of Materials Science and Engineering, Northwestern University, Evanston, IL 60208, USA*
*[4]Department of Materials Science and Engineering, Dankook University, 119 Dandae-ro, Dongnam-gu, Cheonan-si, Chungnam 31116, Republic of Korea*

*[+]These authors contributed equally: Gyu Rac Lee, Thomas Defferriere*

## Abstract

Although strong modulation of interfacial electron concentrations by the relative acidity of surface additives has been suggested, direct observation of corresponding changes in surface conductivity, crucial for understanding the role of local space charge, has been lacking. Here, we introduce a model platform comprising well-aligned mixed ionic-electronic conducting $Pr_{0.2}Ce_{0.8}O_{2-\delta}$ nanowire arrays ($PCO_{NA}$) to show that acidity-modulated heterointerfaces predict electron depletion or accumulation, resulting in tunable electrical properties. We confirm three orders of magnitude increased $PCO_{NA}$ conductivity with basic $Li_2O$ infiltration. Moreover, the relative acidity of the insulating substrate supporting the $PCO_{NA}$ strongly influences its electronic properties as well. This strategy is further validated in purely ionic-conducting nanostructured ceria as well as $PCO_{NA}$. We suggest that observed conductivity changes stem not only from acidity-mediated space charge potentials at heterointerfaces but also from grain boundaries, chemically-modulated by cation in-diffusion. These findings have broad implications for how substrate and surface treatment choices can alter the conductive properties of nanostructured functional oxides.

## Introduction

Heterointerfaces have long been recognized as having a profound influence on the electronic properties of semiconducting materials and their prudent engineering has been vital for the development of a broad variety of microelectronic devices including diodes, transistors, and solar cells[1–3]. This stems from the ability of heterointerfaces to induce band bending, modify charge carrier distributions, and create built-in electric fields that govern the electronic behavior at interfaces[4,5]. The same can be said for ionic materials, where heterointerfaces can lead to extraordinary properties arising from space charge effects with resultant defect redistribution. For example, the formation of two-dimensional electron gases at the heterointerfaces of insulating oxides and the emergence of superconductivity of relevance for quantum applications[6–9] have been, in part, associated with the accumulation of oxygen vacancies acting as electron donors. In more conventional systems, these characteristics can be highly beneficial for enhancing the performance of energy storage and chemical conversion applications, including electrocatalysis, fuel cells, and batteries[10,11].

One representative work by Sata et al. demonstrated marked increases in ionic conductivity with increases in interface density of $CaF_2$ and $BaF_2$ heterostructures[12]. By reducing the heterointerface spacing to below 50 nm, they observed the predicted mesoscopic size effect and revealed that the overlap of space charge regions at nanoscale interfaces can lead to anomalous transport behavior. These findings highlight that precise engineering of the heterointerface enables modulation of the flow of ionic as well as electronic charges. However, studies that successfully exemplify the role of heterointerface space charge in controlling electronic or ionic transport in metal oxides remain limited, except for a few restricted specific compositions[13]. This is due to difficulties in identifying material combinations with appropriate space charge characteristics and the inherent challenge of distinguishing its role from coexisting factors, such as strain[14,15], atomic reconstruction[16], or adsorbate-induced charge localization[17–19] at heterointerfaces. Recently, Steinbach et al. developed a physicochemical model to predict space charge properties at heterointerfaces between various oxides and undoped $SrTiO_3$ single crystals[20]. They were able to relate the experimentally measured space charge potential to defect thermodynamics and the reducibility of oxides. While offering a widely applicable framework, the model depends highly on detailed knowledge of defect formation energies and reducibility of oxides, not always readily available. Moreover, their analysis employs mixed ionic and electronic conducting (MIEC) materials requiring gas-phase equilibrium, necessitating fast kinetics, thereby limiting applicability to high temperatures. Developing simpler material selection criteria for predicting the space charge properties of oxide-based heterointerfaces would go a long way toward providing a practical framework for engineering such composite systems.

A previous study on surface-infiltrated heterostructures by the authors demonstrated that the acidity of binary oxides acts as a sensitive descriptor for predicting the surface oxygen exchange rate ($k_{chem}$) in MIEC electrode[21]. This was attributed to near-surface electron depletion or accumulation, driven by the modulated space charge potential at the heterointerface, which correlated the relative Smith acidity of the surface infiltrated binary oxides relative to that of the host MIEC electrode[21–24]. Building on this insight, the Smith acidity scale, or equivalently the relative work functions of the oxides, serves as a powerful descriptor of electronic and ionic properties induced by local space charges formed at a broader category of heterointerfaces. This serves to provide a rational and useful guideline for materials selection that elucidates the relation between local space charge and conductivity. We furthermore note that the Smith acidity is derived in terms of an oxide's tendency to accept an oxygen anion from a basic oxide during a chemical reaction and calculated using enthalpy changes during such reactions. This framework is consistent with the framework derived by Steinbach et al., as mentioned above.

Although changes in the electrical conductivity of porous $Pr_{0.1}CeO_{0.9}O_{2-\delta}$ layers upon acidic/basic binary oxide infiltration were also assessed, they were modest (10-25%)[21], leading several studies to question the space charge model, despite similar trends being observed in electrode polarization with controlled surface acidity[17–19]. In part, this was based on their inability to detect corresponding changes in the electrical conductivity of their films and little detectable change in the oxidation state of the mixed valent elements in the MIEC surface. We suspect that the smaller changes in conductivity compared to that of $k_{chem}$ were primarily due to the bulk-dominated character of conventional MIEC electrodes[25–27]. Given the nanometer-scale dimensions of space charge regions and the low surface coverage (1.5–3%) achieved during the infiltration process, the influence of acidity-controlled heterointerfaces on overall conductivity was limited[21,28]. These issues therefore necessitate the development of an alternative platform with a much higher surface-to-volume ratio to maximize the influence of surface conduction and examine precisely the impact of heterointerface acidity-mediated space charge on the near-surface electronic properties. Such an approach should serve to unveil the correlation between modified local space charge potentials induced by surface acidity and the corresponding electrical conductivity changes.

In this study, we successfully demonstrate the marked effects of heterointerface acidity on the electronic properties of an optimized nanostructured platform composed of well-aligned $Pr_{0.2}Ce_{0.8}O_{2-\delta}$ nanowire arrays ($PCO_{NA}$). The conductivity of $PCO_{NA}$ is demonstrated to vary by three orders of magnitude over the temperature range of 450-650 °C, following changes in relative Smith acidity of infiltrated binary oxides, ranging from basic ($Li_2O$) to acidic ($SiO_2$). Moreover, serial infiltration with $Li_2O$ following reductions in electrical conductivity induced initially by $SiO_2$, not only recovers, but exceeds the conductivity initially measured by nearly the same factor. We also report, for the first time, and consistent with the above observations, that the relative acidity of insulating substrates ($Al_2O_3$ vs. MgO) supporting $PCO_{NA}$ can also significantly impact its electronic properties. Finally, we note that in polycrystalline materials, space charge potentials at grain boundaries (GBs) must also be accounted for in rationalizing conductivity changes, along with those associated with the heterointerfaces. This is because cation diffusion from substrates or surface infiltrants into polycrystalline thin films can alter their GB space charge potentials, depending on whether in-diffused species act as donors or acceptors at the specific GB lattice sites they occupy. As a result, the overall conduction pathway governing the electronic properties of $PCO_{NA}$ is determined by the acidity-mediated space charge potentials at both the heterointerfaces and GBs, representing clear and strong evidence for the critical role of local space charges in MIEC electrodes.

# Results

## Fabrication and characterization of $PCO_{NA}$ as a model platform

**Fig. 1a** illustrates a cross-sectional view of a $PCO_{NA}$ model platform supported on an insulating substrate to elucidate the role of local space charges induced by modulated surface or substrate acidity on electronic properties. In addition to the bulk conduction ($R_1$), four other possible conduction pathways within $PCO_{NA}$ have been depicted as conduction through substrate heterointerfaces ($R_2$), along GBs ($R_3$), perpendicular to GBs ($R_4$), and surface heterointerfaces with infiltrants ($R_5$). Because space charge potentials at heterointerfaces and GBs are affected by the relative acidity of binary oxides and the in-diffusion of cations along GBs sourced from the binary oxides, all these conduction pathways need to be considered. Notably, all other conduction pathways are always in series with conduction perpendicular to the GBs ($R_4$), so the overall equivalent circuit can be described as shown in **Fig. 1b**.

To achieve a high surface-to-volume ratio to enhance the relative ratio of surface to bulk conduction, $PCO_{NA}$ was fabricated using a solvent-assisted nanotransfer printing method (S-nTP) (**Fig. 1c**), previously reported[29–31]. This morphology offers readily accessible open structures for the conformal formation of interfaces with both substrates and infiltrants. As shown in the scanning electron microscopy (SEM) images (**Fig. 1d**), $PCO_{NA}$ with widths of 50 nm, pitch of 150 nm, and thickness of 50 nm were fabricated on insulating substrates. Due to the shadowing effect caused by oblique angle deposition[32], $Pr_{0.2}Ce_{0.8}O_{2-\delta}$ (PCO20) was deposited only on one side of the PMMA, resulting in an unconventional droplet-shaped structure with a tail-like cross-section (see inset). This nanoscale morphology contributes to an increased surface-to-volume ratio, intensifying the influence of surfaces properties compared to that of the bulk, and provides a large number of parallel conductive nanowire channels (estimated to be 50,000) for precise analysis of the in-plane electrical conductivity. The concentration of Pr (20 at%) in the fabricated $PCO_{NA}$ was identical to that of the target source as confirmed by inductively-coupled plasma mass spectrometry (ICP-MS) (**Supplementary Table 1**). The accompanying transmission electron microscopy (TEM) image shows that the PCO20 nanowires are composed of nanoscale grains with diameters ranging from a few up to 20 nm (**Fig. 1e**), and that these nanosized grains result in a rough surface, which amplifies the impact of the surface on properties. While PCO20 is shown to be polycrystalline, most of the crystal planes are observed to consist of $CeO_2$ (111) planes with d-spacings of 0.31 nm. X-ray diffraction (XRD) spectra also assign the main peak to the (111) plane, although other oriented planes also exist (**Fig. 1f**). Furthermore, the identical positions of peaks appearing in fabricated $PCO_{NA}$ compared to as-transferred samples imply that strain effects are negligible[14,33]. X-ray photoelectron spectroscopy (XPS) shows only peaks associated with Ce/Pr, with no evidence of other species such as Sr and Cr which would cause degradation as shown in **Supplementary Fig. 1**. Finally, Au current collectors were sequentially deposited on both ends of the $PCO_{NA}$ for evaluating in-plane conductivity. The distance between the two current collectors and the width of the model platform were fixed to 1 mm and 1 cm, respectively.

**Surface-infiltrated oxide acidity effects on $PCO_{NA}$ electronic properties**

To investigate the effects of binary oxide acidity on the electrical conductivity of $PCO_{NA}$, two $PCO_{NA}$ specimens with different binary oxides infiltrated onto their surfaces were prepared and deposited on single crystal sapphire ($Al_2O_3$) substrates. The in-plane conductivity of uninfiltrated $PCO_{NA}$ as well as $PCO_{NA}$ infiltrated by $SiO_2$ (acidic) and by $Li_2O$ (basic) were measured utilizing AC impedance spectroscopy, respectively. Additionally, to explore the potential for recovering degraded conductivity, $PCO_{NA}$ with serial infiltration of Si-species followed by Li-species was also prepared and evaluated[24].

The primary objective of this study was to monitor conductivity changes induced by heterointerfaces through the surface infiltration of binary oxides with different acidity, rather than the initial focus on the oxygen exchange rate as reported previously[21,34,35]. Hence, we attempted to achieve full $PCO_{NA}$ coverage by using a concentrated solution (0.2 M) to maximize coverage and as shown in SEM images in **Supplementary Fig. 2,** the coverage is complete. It is important to note that the electrical conductivities of the insulating binary oxide infiltrants used in this study are negligible compared to that of the PCO nanowires (**Supplementary Table 2**). The measured impedance spectra are therefore assumed to solely represent $PCO_{NA}$ electrical pathways. AC impedance spectra were obtained between 450 – 650 °C to extract the in-plane conductivity of both the uninfiltrated and infiltrated samples (**Supplementary Fig. 3**).

As illustrated in **Fig. 2a** and **2b**, Li-infiltration resulted in nearly three orders of magnitude decrease of resistance (see inset), while Si-infiltration caused only a slight increase in resistance, both relative to the uninfiltrated $PCO_{NA}$. Arrhenius plots of in-plane conductivity under each infiltration condition, along with the corresponding activation energy ($E_a$) values, are summarized in **Fig. 2c**. Consistent with the impedance plots, Li-infiltrated $PCO_{NA}$ exhibited approximately three orders of magnitude higher in-plane conductivity compared to uninfiltrated $PCO_{NA}$ over the full temperature range, with a same nominal $E_a$ of 0.85±0.023 eV.

However, only 1.5 times lower conductivity was measured in Si-infiltrated $PCO_{NA}$ relative to the uninfiltrated reference with similar $E_a$ (0.89±0.025 eV). A similar trend was also observed in serial infiltration with Li-species following Si infiltration, aimed at recovering conductivity, as shown in **Fig. 2b** (see also inset). Subsequent Li infiltration on Si-infiltrated $PCO_{NA}$ not only restored, but exceeded the initial conductivity by 100-fold with a reduced $E_a$ of 0.63±0.065 eV. Although serial Li infiltration did not achieve as much improvement in conductance as sole Li infiltration, given the pre-coating effects of $SiO_2$, it once again verified the strong positive impact of basicity on already degraded electrodes.

**Insulating substrate acidity effects on $PCO_{NA}$ electronic properties**

We next explored the impact of the choice of insulating substrate on the electronic properties of our $PCO_{NA}$. This is not only of relevance to further validate the observed trends of heterointerface acidity on the electrical conductivity of $PCO_{NA}$, but also considering that diverse insulating substrates, such as $Al_2O_3$ or MgO, are commonly used to study the electrical properties of deposited oxide thin films. Generally, one employs such substrates to support controlled film growth while minimizing the potential for short-circuiting of the supported films through the substrate. The potential impact of their interfacial properties on the overlaying films' conductivity, however, is generally ignored based on the assumption that transport along such interfaces would have an insignificant impact compared to transport through the bulk ($R_1$). It is noteworthy that most of these insulating substrates each possess their own distinct relative Smith acidities, and thereby work functions, which could presumably result in inducing differing space charge phenomena at these interfaces[36]. $PCO_{NA}$ were fabricated on both $Al_2O_3$ and MgO substrates (**Supplementary Fig. 4**), expected to exhibit acidic and basic characteristics respectively relative to PCO20. This enabled us to assess any alterations in the in-plane conductivity induced at the heterointerface between the substrates and $PCO_{NA}$.

The AC impedance spectra of $PCO_{NA}$ fabricated on the two different substrates were obtained for temperatures between 450 – 650 °C as shown in **Supplementary Fig. 3 and 5**. By comparison of these two impedance spectra at the same measurement temperature of 650 °C (**Fig. 3a**), it becomes immediately obvious that $PCO_{NA}$ fabricated on $Al_2O_3$ is over an order of magnitude more resistive than the same structure fabricated on MgO. Arrhenius plots of in-plane conductivity are shown in **Fig. 3b** along with a listing of their respective $E_a$ values. Consistent with the impedance plots, $PCO_{NA}$ on the MgO substrate exhibited approximately 10 times higher in-plane conductivity compared to $PCO_{NA}$ on the $Al_2O_3$ substrate over the full temperature range, with a nominally lower $E_a$ of 0.78±0.037 eV vs. 0.85±0.007 eV. Since all other conditions were identical except for the choice of substrate, it is reasonable to assume that the differences in conductivity between these two samples can be attributed to the relative acidity of the substrates. With this insight, it can be argued that the selection of a suitable substrate must be considered before depositing nanostructured semiconductors or MIECs to avoid such heterointerface acidity effects. Indeed, we suspect that many reports of thin film oxide conductivity have been impacted by the relative substrate acidity of the supporting substrate and thus need to be reevaluated or appropriately normalized.

# Discussions

## Acidity-mediated space charge potential at heterointerfaces and GBs

Based on the relative acidity of infiltrants or substrates relative to PCO, adjacent acidic/basic species were expected to induce electron depletion/accumulation at the heterointerfaces with $PCO_{NA}$, thereby leading to increased/decreased resistance along both the conduction pathways through substrate ($R_2$) and surface ($R_5$) heterointerfaces. As we have shown in **Fig. 2c** and **3b**, elements with higher acidity relative to PCO ($Al_2O_3$ and $SiO_2$) resulted in decreases in conductivity, while elements with lower acidity relative to PCO (MgO and $Li_2O$) resulted in increases in conductivity, consistent with our expectation. However, several details need to be explained in more detail to fully appreciate the observed trends:

I) The relative magnitude of change in the conductivity of $PCO_{NA}$ driven by surface acidity modulation with basic and acidic elements compared to uninfiltrated samples differed significantly as illustrated in **Fig. 2c** and **3b**. This divergence might originate from the distinct conduction pathways determined by space charge potentials at the heterointerfaces. For example, in the case of acidic treatments, the higher relative acidity compared to that of PCO is expected to generate a more resistive conduction pathway through the surface/substrate heterointerfaces ($R_5$/$R_2$), leading to the expectation that conduction occurs primarily through the bulk ($R_1$) (**Fig. 4a** and **4c**). On the other hand, the treatment with basic elements will result in carrier enhancement due to electron accumulation and thereby enhanced conduction through the surface/substrate heterointerface ($R_5$/$R_2$) (**Fig. 4b**). Interestingly, Li-infiltration had a much larger impact (three orders of magnitude at 450 °C) on the $PCO_{NA}$ conductivity than the MgO substrate (factor of ~10 at 450 °C), although both have a higher basicity than PCO. This can be related to (i) the very different natures of the two interfaces (the unusual droplet-shape of the top rough surface of $PCO_{NA}$ leads to a much higher interface area than at the planar substrate interface) and (ii) the higher relative basicity of $Li_2O$ is expected to result in higher electron accumulation and therefore higher conductivity compared to MgO.

II) In the case of acidic treatment with $SiO_2$ infiltration or use of $Al_2O_3$ substrate, one would expect, as mentioned above, the conductivity to solely conduct through the bulk ($R_1$), and therefore for it to reach similar values to that of the bulk PCO20 reference. However, the conductivity and its $E_a$ for $PCO_{NA}$ are reduced in both cases compared to bulk PCO20 (1 eV), which indicates that conduction is impacted by pathways in addition to the bulk conduction ($R_1$). Here one must consider how grain boundaries (GBs), as well, can potentially serve as either short-circuiting pathways ($R_3$) or serve to block transport ($R_4$), as illustrated in **Fig. 1a** and **1b**. While we do not report any direct measurements of GB characteristics in this study, based on their expected behavior, we can explain the above observations. In the following, we therefore briefly review key relevant findings describing the roles of GBs in impacting charge transport in ceria-based systems and identify how small polaron hopping mobility, characteristic of electron transport in PCO, can result in GB phenomena contrary to normal expectations. For those interested, more detailed discussions of these issues may be found in **Supplementary Note 2**.

GBs in ceria systems[37,38], in general, exhibit a positive space charge potential associated with a positive GB core charge, causing a depletion of oxygen vacancies and accumulation of electrons in the adjacent space charge zones[39]. In nominally undoped $CeO_2$, the accumulation of electrons has been shown to lead to electronically conducting layers that bypass bulk ionic transport, thereby controlling the overall conductivity[40–42]. On the other hand, as mentioned above, this same positive GB space charge potential in acceptor-doped $CeO_2$ (e.g. Gd doped $CeO_2$ - GDC), an oxygen ion conductor, can lead to orders of magnitude reduction in oxygen ion conduction due to oxygen vacancy depletion at the GBs[37,43]. As recently demonstrated by the authors, this space charge potential can be tuned by in-diffusion of substrate elements (i.e. Mg and Al) into GDC thin films at intermediate temperatures (700-900 °C)[43], as described in **Supplementary Note 2**. We suspect correspondingly significant Al up-diffusion occurred along GBs in $PCO_{NA}$ samples grown on alumina substrates annealed at 700 °C (**Fig. 4c, lower image**), compared to those annealed at 400 °C (**Fig. 4a**), leading to an increase of the positive space charge potential at the GBs (**Fig. 4c**). While conduction along the GBs ($R_3$), in principle, should give enhanced conduction due to accumulation of electrons around GBs, that transport path nevertheless would have to pass through a mobility blockage related to the small polaron hopping controlling electronic transport in PCO connected with the path perpendicular to GBs ($R_4$), as discussed in **Supplementary Note 2** and displayed in **Supplementary Fig. 6** and **7**. This results in reduced $E_a$, attributed to the enhanced space charge potential at the GBs, yet without a corresponding increase in conductivity. In fact, it is even lower than that of the $PCO_{NA}$ annealed at 400 °C (**Supplementary Fig. 8**), where Al up-diffusion is expected to be kinetically restricted and therefore the space charge potentials at the grain boundaries are lower. The lower conductivity is related to the inability of electrons to hop from occupied $Pr^{3+}$ sites to adjacent unoccupied $Pr^{4+}$ sites under conditions of high electron accumulation at the GBs. XRD spectra showed no difference in peak position and shape of PCONA annealed at different temperatures suggesting that no differences in strain were present (**Supplementary Fig. 9**). On the other hand, in the case of the nanowires sitting on MgO, transport across GBs ($R_4$) become less resistive due to the up-diffusion of Mg at the GBs, lowering the space charge potential, allowing for the measurement of the un-impeded conductivity enhancement parallel to the substrate heterointerface ($R_2$) (**Fig. 4b**). Thus, the similarities in $E_a$ between the $PCO_{NA}$ treated with acids and bases, but different conductivity trends, originate from the presence of positive space charge potential at different dominant conduction pathways.

Interestingly, opposite trends were observed in a purely ionic conducting ceria system, 3 mol% Gd-doped ceria (GDC3), prepared via the same method as $PCO_{NA}$ (**Supplementary Note 3**). Here the basic treatment with MgO as a substrate leads to a total decrease in conductivity in contrast to $Al_2O_3$. This observation aligns very well with the space charge model, as the majority of carriers in this case are positively charged oxygen vacancies ($V_O^{\cdot\cdot}$), and therefore the space charges present at the heterointerface are expected to impact the carrier in the opposite direction to the electronic carriers in $PCO_{NA}$. The activation energies measured in the GDC3 nanowire arrays are characteristic of grain boundary-controlled resistance and are clearly influenced by the substrate heterointerfaces, similar to the case of the $PCO_{NA}$ (**Supplementary Fig. 10**). Therefore, the Smith acidity scale is a universally applicable descriptor for predicting the influence of heterointerface space charge effects not only on electronic but also ionic carriers.

III) Finally, the relatively moderate enhancement achieved in sequential infiltration with Li-species following Si (**Fig 2c**), compared to solely Li-infiltrated $PCO_{NA}$, can be attributed to the effective Smith acidity of Li-Si compounds[44]. As evidenced by the XPS spectra (**Supplementary Fig. 11**), Li-species react with $SiO_2$ at high temperatures, forming intermediate lithium silicate composites ($Li_2SiO_3$ or $Li_4SiO_4$). The acidity of these composites should be located between the Smith acidity scale of $Li_2O$ and $SiO_2$, which can be expressed as:

$$a_{\mathrm{Li_2SiO_3}} = \frac{2a_{\mathrm{Li_2O}} + a_{\mathrm{SiO_2}}}{3} \; or \; a_{\mathrm{Li_4SiO_4}} = \frac{4a_{\mathrm{Li_2O}} + a_{\mathrm{SiO_2}}}{5}$$

Due to the higher basicity of $Li_2O$ and its dominant influence, lithium silicate exhibits strong basicity, in contrast to $SiO_2$, and plays a role in accumulating electrons at the top heterointerface of $PCO_{NA}$, although obviously not as high as $Li_2O$ alone. This leads to the formation of a dominant conduction pathway through the surface heterointerface ($R_5$). This change in effective Smith acidity leads to a 100-fold improvement in conductivity.

We suggest that frameworks driven by the Smith acidity scale enable us to predict the space charge properties of oxide-based heterointerfaces. One can therefore expect application of Smith acidity criteria in predicting heterointerface properties to have a very broad impact given that they can be readily applied not only to MIECs, demonstrated by Steinbach et al., but also to semiconducting and even insulating oxides. This makes the approach suitable for application to a much wider temperature range of devices even for microelectronics operating at room temperature. Furthermore, the proposed framework achieved both enhancement and suppression of space charge potential depending on the relative surface acidity. By demonstration and modeling this approach with polycrystalline $PCO_{NA}$ rather than single-crystal samples, it was possible to successfully account for charge transport behavior not only across but also along the heterointerfaces and GBs. This served to provide more realistic insights into acidity-mediated space charge engineered conductivity, with significant implications for practical nanostructured thin film applications.

## Summary and outlook

In summary, we have investigated how the $PCO_{NA}$ model platform enables direct observation of orders of magnitude conductivity changes induced at the respective heterointerfaces and GBs, confirming the influence and significance of surface acidity-mediated local space charge potentials. These correlations were firmly established by examining the electronic properties of $PCO_{NA}$ with different infiltrants, insulating substrates, and annealing temperatures. Notably, a single Li-infiltration led to a remarkable three orders of magnitude increase in conductivity. Moreover, our findings challenge the assumption in many studies that insulating substrates do not affect electrical measurements, as their relative acidity can significantly impact measured conductivity at the film substrate interface and upon higher temperature anneals, impact film GB properties as well. Furthermore, our study highlights how the in-diffusion of heterointerface elements at film GBs can also markedly modify the GB transport properties, which needs to be taken into account to fully understand the total changes in conductivity of such nanocrystalline thin films. These features play a direct and critical role in determining the electronic properties of nanostructured functional oxides. This strongly supports the concept of the Smith acidity scale as a predictive descriptor for space charge behavior at heterointerfaces in functional and electrocatalytically active oxide systems. Our understanding of the possible role of supporting substrate in modulating the electronic properties of thin oxide films opens up new opportunities for the modulation of thin film properties without altering their defect chemistry, oxygen stoichiometry, or microstructure.

# Methods

### Preparation of $Pr_{0.2}Ce_{0.8}O_{2-\delta}$ (PCO20) target source for pulsed laser deposition (PLD)

The PCO20 oxide target was synthesized through a sol-gel method, involving dissolving cerium and praseodymium nitrates with chelating agents in water, adjusting the pH to 9.5, and heating at 80 °C to induce gelation. The gel was dried, fired at 450 °C, calcined at 750 °C to produce PCO20 powder, which was then pelletized and sintered at 1400 °C to form the PLD target.

### Fabrication of $Pr_{0.2}Ce_{0.8}O_{2-\delta}$ nanowire arrays ($PCO_{NA}$) on insulating substrates

$PCO_{NA}$ were fabricated using a combination of the solvent-assisted nanotransfer printing (S-nTP) method and PLD, as illustrated in **Fig. 1a**. Initially, a Si master mold with line-patterned features (line width: 50 nm, line pitch: 150 nm) was prepared via KrF photolithography and reactive ion etching. To facilitate easy detachment of the polymer transfer medium, a layer of polydimethylsiloxane (PDMS, Polymer Source Inc.) was applied to the Si master mold. Subsequently, a solution of polymethylmethacrylate (PMMA, Sigma-Aldrich Inc.) dissolved in a mixed solvent of toluene, acetone, and heptane was spin-cast onto the pre-treated Si master mold to create the polymer transfer medium. The reason for mixing those three solvents is to optimize the surface energy of the layer with respect to the substrate. The PMMA transfer medium was then peeled off from the Si master mold using a polyimide adhesive tape (PI, 3M Inc.), resulting in a reverse morphology of the Si master mold shape. $Pr_{0.2}Ce_{0.8}O_{2-\delta}$ was deposited using PLD with oblique-angle of 80° to guide deposition selectively onto the sections that protrude from the PMMA transfer medium. PLD (Coherent COMPex Pro 205) was performed using a 248 nm KrF excimer laser operating at an energy level of 300 mJ under a frequency of 10 Hz at room temperature. Following deposition, discrete $PCO_{NA}$ were obtained. $PCO_{NA}$ were transfer-printed onto target substrates ($Al_2O_3$ and MgO, in this study) and the PMMA was removed via washing with toluene or acetone. Due to the limitations imposed by the PMMA transfer medium, high-temperature deposition for improving crystallinity during PLD was not feasible. Thus, the resulting $PCO_{NA}$ on the target substrates were annealed at 700 °C under ambient atmosphere for 2 hours.

### Physicochemical characterization of materials

The physical and chemical characterization of $PCO_{NA}$ involved a range of techniques including scanning electron microscopy (SEM), transmission electron microscopy (TEM), X-ray diffraction (XRD), X-ray photoelectron spectroscopy (XPS), and inductively coupled plasma mass spectrometry (ICP-MS). SEM analysis was performed using a Hitachi S-4800 with an acceleration voltage of 10 kV, while TEM imaging was conducted with FEI Tecnai G2 F30 S-Twin operating at an acceleration voltage of 300 kV. High-angle annular dark-field scanning TEM (HAADF-STEM), selected area electron diffraction (SAED) pattern, and energy dispersive X-ray spectrometer (EDX) mapping were obtained using a TEM (JEOL, JEM-2100F HR) operated at an acceleration voltage of 200 kV. Crystal information of the $PCO_{NA}$ was determined through XRD (RIGAKU, SmartLab) conducted in θ–2θ scan mode with a Cu Kα1 incident beam. XPS (Thermo VG Scientific, K-Alpha) measurements provided insights into the chemical compositions, oxidation states and charge transfer evidence, with the C 1*s* peak at 284.8 eV serving as the reference for binding energy calibration. ICP-MS (Agilent, ICP-MS 7700S) was employed to analyze the chemical composition of $PCO_{NA}$, with experiments repeated at least five times to ensure accuracy.

**Electrical conductivity measurements**

The geometric factors associated with the nanowire-formed electrically conductive channels assured that their overall resistance was significantly larger than that of the electrode contact resistance. Thus, the electrical properties measured could be predominantly attributed to the $PCO_{NA}$. Electrical conductivity assessments of the $PCO_{NA}$ were conducted using a two-probe electrode system. AC impedance spectroscopy was carried out using a Solartron 1255 HF frequency response analyzer and EG&G PAR potentiostat across a frequency range of 0.1 Hz to 100,000 Hz, within a temperature range of 450 – 650 °C (at intervals of 50 °C), under varying oxygen partial pressure (ranging from 0.1 to 1 atm). AC impedance spectroscopy was applied to ensure the ability to isolate bulk (high frequency contribution) from electrode (low frequency) contributions and in-plane conductivity measurements were made while lowering the temperature at each annealing temperature. All the impedance spectra were fit to an equivalent RC circuit in which the low frequency intercept of the high frequency semicircle corresponds to the resistance of the $PCO_{NA}$ at each measurement temperature. The experimental setup involved placing the samples in an alumina tube, with a gold current collector linked to a gold wire, while gas mixtures of oxygen and nitrogen were regulated through digital mass flow controllers. In-plane conductivity was monitored by recording voltage measurements every 0.5 s until equilibrium was reached using AC measurements. Initial conductivities of $PCO_{NA}$ without infiltration were evaluated to establish reference values.

**Surface infiltration of binary oxides on $PCO_{NA}$**

The infiltration method involved a simple drop and drying process on $PCO_{NA}$ with ethanol-based solutions of tetraethylorthosilicate (TEOS) and $Li(NO_3)_3$. Solutions containing Li nitrate and TEOS at a concentration of 0.2 M were prepared for the infiltration process. Initial conductivities of $PCO_{NA}$ without infiltration were evaluated to establish reference values. Infiltration was performed with the sample located in its original cell setup, ensuring complete solution coverage and drying of 50 μL, thereby enhancing infiltration reliability. These precursors decompose and convert to binary oxides ($SiO_2$ and $Li_2O$, respectively) at temperatures exceeding 600 °C under air atmosphere[23]. Thus, $PCO_{NA}$ was annealed under air conditions at 650 °C and measurements initiated after stabilization of the conductivity was achieved.

# Figures

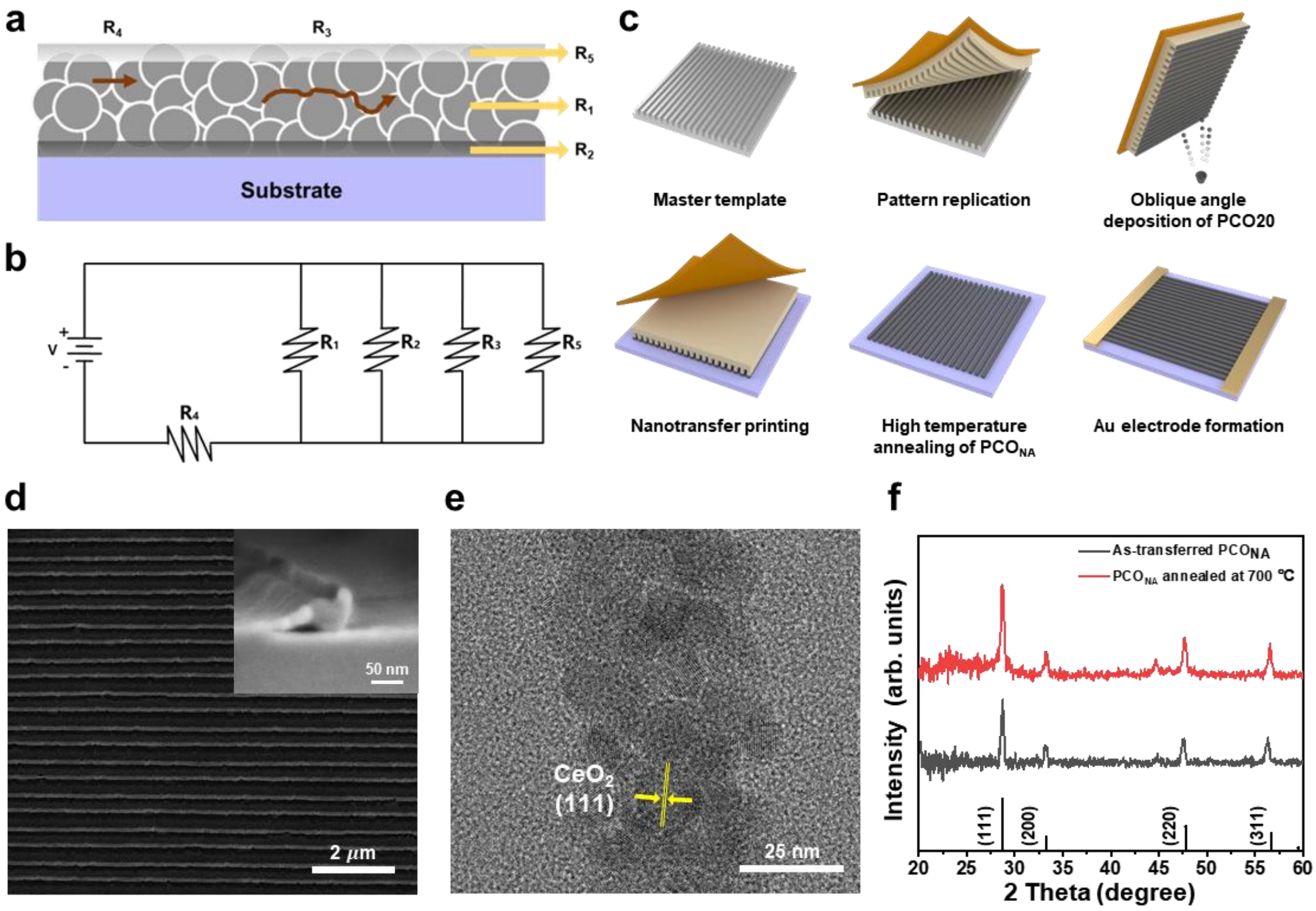

**Fig. 1 Fabrication and characterization of PCO20 nanowire arrays ($PCO_{NA}$) as model platform for analyzing effects of heterointerface acidity on electronic properties.**
**a** Schematic cross-sectional illustration of several conduction pathways in PCO nanowires. **b** Corresponding equivalent circuit representing the interconnection of the conduction pathways illustrated in **a**. **c** Fabrication process of $PCO_{NA}$ model platform through solvent-assisted nanotransfer printing method with pulsed laser deposition and high-temperature annealing. **d** SEM image of fabricated $PCO_{NA}$ (inset image: Cross-sectional SEM image of fabricated $PCO_{NA}$) and **e** TEM image of fabricated PCO nanowire. **f** XRD spectra of as-transferred $PCO_{NA}$ and fabricated $PCO_{NA}$ after high-temperature annealing at 700 °C. Black vertical lines at bottom of figure indicate the dominant facets of the cerium oxide crystal structure.

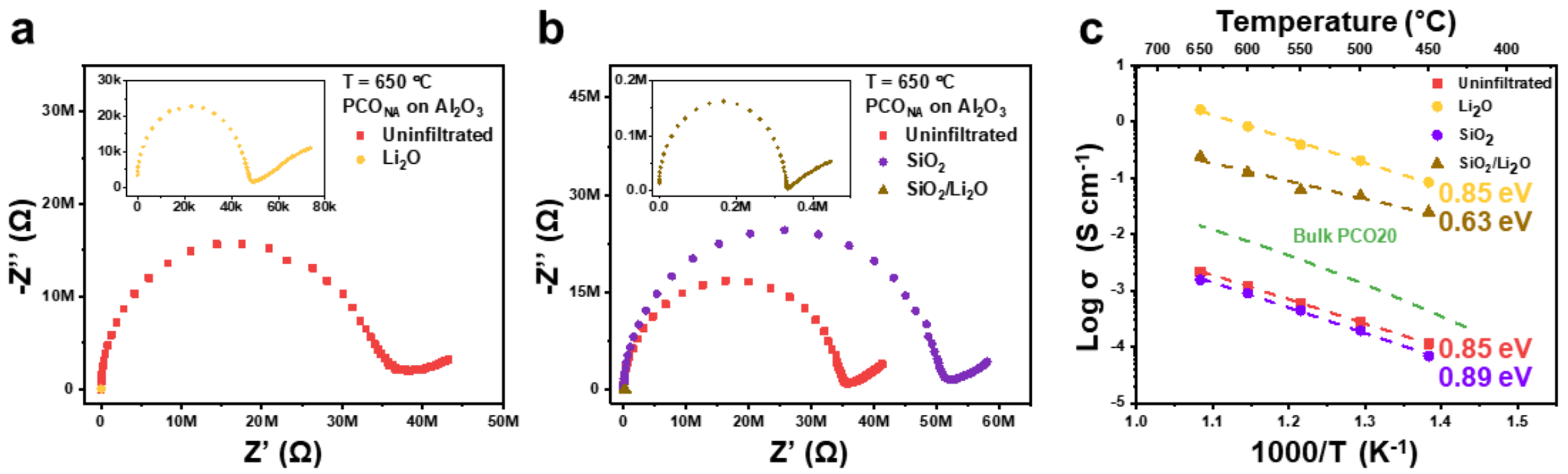

**Fig. 2 Influence of infiltrated binary oxide acidity on $PCO_{NA}$ electronic properties.**
**a** Impedance spectra of pristine $PCO_{NA}$ followed by Li-infiltration at 650 °C (see inset for reduced resistance). **b** Impedance spectra of pristine $PCO_{NA}$ followed by Si-infiltration and serial Li-infiltration for recovering conductance (see inset for reduced resistance) at 650 °C. $Al_2O_3$ used as substrate in both cases. **c** Arrhenius plots of in-plane conductivity of pristine $PCO_{NA}$ followed by Li-, Si- and serial Li-infiltrations along with their respective activation energies, respectively. The green dashed line indicates the conductivity behavior of a bulk PCO20 film calculated based on defect chemistry modelling.

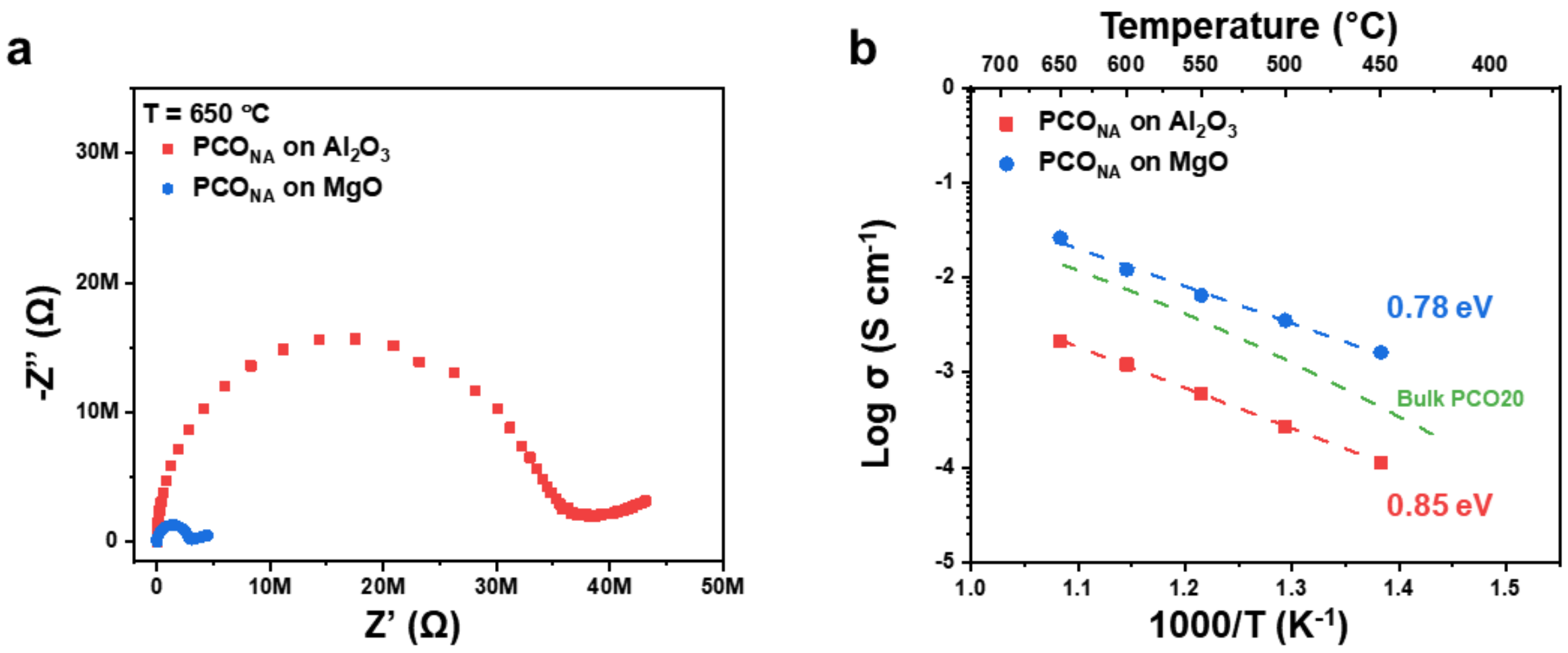

**Fig. 3 Influence of insulating substrate acidity on PCO$_{NA}$ electronic properties.**

**a** Comparison of impedance spectra of PCO$_{NA}$ fabricated on $Al_2O_3$ and MgO substrates measured at 650 °C. **b** Arrhenius plots of in-plane conductivity of PCO$_{NA}$ fabricated on $Al_2O_3$ and MgO substrates along with their respective activation energies. The green dashed line indicates the conductivity behavior of a bulk PCO20 film calculated based on defect chemistry modelling.

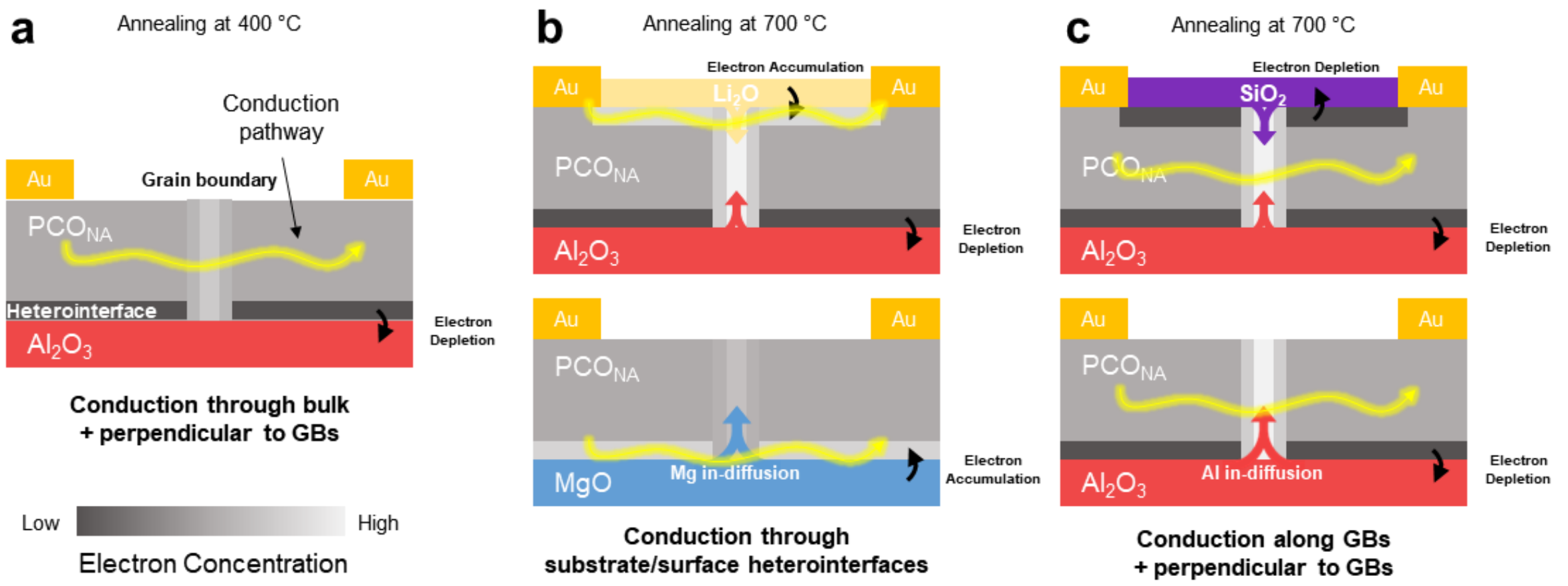

**Fig. 4 Schematic illustration of heterointerface acidity-mediated space charge potentials at heterointerfaces and GBs of $PCO_{NA}$, and impact of electron redistribution**

**a** $PCO_{NA}$ annealed at 400 °C. Conduction mainly occurs through bulk ($R_1$) and perpendicular to GBs ($R_4$). Substrate heterointerface shown in black corresponds to electron depletion due to the relative acidity of $Al_2O_3$. Slightly brighter GB is attributed to a built-in space charge potential (positive core charge) leading to some electron accumulation. For high-temperature annealing, cation diffusion into GBs must be considered. Diffusion of cations up to GBs leads to GB electron accumulation/depletion, described in white/gray (see **b** and **c**). **b** Li-infiltrated $PCO_{NA}$ (upper image) and $PCO_{NA}$ fabricated on MgO substrate (lower image) annealed at 700 °C. Conduction mainly occurs through substrate ($R_2$) (upper image) or surface ($R_5$) (lower image) heterointerfaces. Substrate/surface heterointerface shown in light grey corresponds to electron accumulation due to the relative basicity of $Li_2O$ and MgO. **c** Si-infiltrated $PCO_{NA}$ (upper image) and $PCO_{NA}$ fabricated on $Al_2O_3$ substrate (lower image) annealed at 700 °C. Conduction mainly occurs along GBs ($R_3$) and perpendicular to GBs ($R_4$). Substrate/surface heterointerface shown in black corresponds to electron depletion due to the relative acidity of $SiO_2$ and $Al_2O_3$. The scale of electron concentration is represented through the color bar at the bottom left.

# References

1. Wang, H. *et al.* Semiconductor heterojunction photocatalysts: design, construction, and photocatalytic performances. *Chem. Soc. Rev.* **43**, 5234–5244 (2014).

2. Gbadamasi, S. *et al.* Interface chemistry of two-dimensional heterostructures – fundamentals to applications. *Chem. Soc. Rev.* **50**, 4684–4729 (2021).

3. Duan, T. *et al.* Chiral-structured heterointerfaces enable durable perovskite solar cells. *Science* **384**, 878–884 (2024).

4. Bediako, D. K. *et al.* Heterointerface effects in the electrointercalation of van der Waals heterostructures. *Nature* **558**, 425–429 (2018).

5. Tan, S. *et al.* Stability-limiting heterointerfaces of perovskite photovoltaics. *Nature* **605**, 268–273 (2022).

6. Thiel, S., Hammerl, G., Schmehl, A., Schneider, C. W. & Mannhart, J. Tunable quasi-two-dimensional electron gases in oxide heterostructures. *Science* **313**, 1942–1945 (2006).

7. Ohtomo, A. & Hwang, H. Y. A high-mobility electron gas at the $LaAlO_3/SrTiO_3$ heterointerface. *Nature* **427**, 423–426 (2004).

8. Liu, C. *et al.* Two-dimensional superconductivity and anisotropic transport at $KTaO_3$ (111) interfaces. *Science* **371**, 716–721 (2021).

9. Iannaccone, G., Bonaccorso, F., Colombo, L. & Fiori, G. Quantum engineering of transistors based on 2D materials heterostructures. *Nat. Nanotechnol.* **13**, 183–191 (2018).

10. Gao, Z., Mogni, L. V., Miller, E. C., Railsback, J. G. & Barnett, S. A. A perspective on low-temperature solid oxide fuel cells. *Energy Environ. Sci.* **9**, 1602–1644 (2016).

11. Zheng, J. *et al.* Design principles for heterointerfacial alloying kinetics at metallic anodes in rechargeable batteries. *Sci. Adv.* **8**, eabq6321 (2022).

12. Sata, N., Eberman, K., Eberl, K. & Maier, J. Mesoscopic fast ion conduction in nanometre-scale planar heterostructures. *Nature* **408**, 946–949 (2000).

13. Liang, C. C. Conduction Characteristics of the Lithium Iodide-Aluminum Oxide Solid Electrolytes. *J. Electrochem. Soc.* **120**, 1289 (1973).

14. Harrington, G. F. *et al.* Tailoring Nonstoichiometry and Mixed Ionic Electronic Conductivity in $Pr_{0.1}Ce_{0.9}O_{2-\delta}/SrTiO_3$ Heterostructures. *ACS Appl. Mater. Interfaces* **11**, 34841–34853 (2019).

15. Schichtel, N. *et al.* Elastic strain at interfaces and its influence on ionic conductivity in nanoscaled solid electrolyte thin films–theoretical considerations and experimental studies. *Phys. Chem. Chem. Phys.* **11**, 3043–3048 (2009).

16. Garcia-Barriocanal, J. *et al.* Colossal ionic conductivity at interfaces of epitaxial $ZrO_2$:$Y_2O_3$/$SrTiO_3$ heterostructures. *Science* **321**, 676–680 (2008).

17. Siebenhofer, M. *et al.* Engineering surface dipoles on mixed conducting oxides with ultra-thin oxide decoration layers. *Nat. Commun.* **15**, 1–10 (2024).

18. Riedl, C. *et al.* Surface Decorations on Mixed Ionic and Electronic Conductors: Effects on Surface Potential, Defects, and the Oxygen Exchange Kinetics. *ACS Appl. Mater. Interfaces* **15**, 26787–26798 (2023).

19. Schwenkel, D. M., De Souza, R. A. & Harrington, G. F. Understanding the role of acidity on the surface exchange reaction in mixed conductors: what is the effect of surface hydration? *J Mater. Chem. A* **13**, 8541–8548 (2025).

20. Steinbach, C. *et al.* Space Charges at $SrTiO_3$|Mixed Ionic and Electronic Conducting Oxide Heterojunctions and Their Relation to Defect Chemistry. *ACS Appl. Mater. Interfaces* **17**, 17543–17557 (2025).

21. Nicollet, C. *et al.* Acidity of surface-infiltrated binary oxides as a sensitive descriptor of oxygen exchange kinetics in mixed conducting oxides. *Nat. Catal.* **3**, 913–920 (2020).

22. Staerz, A., Seo, H. G., Defferriere, T. & Tuller, H. L. Silica: ubiquitous poison of metal oxide interfaces. *J Mater. Chem. A* **10**, 2618–2636 (2022).

23. Seo, H. G. *et al.* Reactivation of chromia poisoned oxygen exchange kinetics in mixed conducting solid oxide fuel cell electrodes by serial infiltration of lithia. *Energy Environ. Sci.* **15**, 4038–4047 (2022).

24. Seo, H. G. *et al.* Tuning Surface Acidity of Mixed Conducting Electrodes: Recovery of Si-Induced Degradation of Oxygen Exchange Rate and Area Specific Resistance. *Adv. Mater.* **35**, 2208182 (2023).

25. Ding, D., Li, X., Lai, S. Y., Gerdes, K. & Liu, M. Enhancing SOFC cathode performance by surface modification through infiltration. *Energy Environ. Sci.* **7**, 552–575 (2014).

26. Kim, S. *et al.* Microstructure tailoring of solid oxide electrolysis cell air electrode to boost performance and long-term durability. *Chem. Eng. J.* **410**, 128318 (2021).

27. Chen, G. *et al.* Roadmap for Sustainable Mixed Ionic-Electronic Conducting Membranes. *Adv. Funct. Mater.* **32**, 2105702 (2022).

28. Crumlin, E. J. *et al.* Surface strontium enrichment on highly active perovskites for oxygen electrocatalysis in solid oxide fuel cells. *Energy Environ. Sci.* **5**, 6081–6088 (2012).

29. Jeong, J. W. *et al.* High-resolution nanotransfer printing applicable to diverse surfaces via interface-targeted adhesion switching. *Nat. Commun.* **5**, 1–12 (2014).

30. Lee, G. R. *et al.* Efficient and sustainable water electrolysis achieved by excess electron reservoir enabling charge replenishment to catalysts. *Nat. Commun.* **14**, 1–12 (2023).

31. Lee, G. R. *et al.* Unraveling oxygen vacancy-driven catalytic selectivity and hot electron generation on heterointerfaces using nanostructured platform. *Nat. Commun.* **16**, 1–12 (2025).

32. Kim, Y. J. *et al.* Highly efficient oxygen evolution reaction via facile bubble transport realized by three-dimensionally stack-printed catalysts. *Nat. Commun.* **11**, 1–11 (2020).

33. Zhang, S., Fang, Z., Chi, M. & Perry, N. H. Facile Interfacial Reduction Suppresses Redox Chemical Expansion and Promotes the Polaronic to Ionic Transition in Mixed Conducting $(Pr,Ce)O_{2-\delta}$ Nanoparticles. *ACS Appl. Mater. Interfaces* **17**, 880–898 (2024).

34. Seo, H. G., Kim, H., Jung, W. C. & Tuller, H. L. Reversal of chronic surface degradation of $Sr(Ti,Fe)O_3$ perovskite-based fuel cell cathodes by surface acid/base engineering. *Appl. Catal. B–Environ. Energy* **355**, 124172 (2024).

35. Kim, H., Seo, H. G., Ahn, S., Tuller, H. L. & Jung, W. C. Unveiling critical role of metal oxide infiltration in controlling the surface oxygen exchange activity and polarization of $SrTi_{1-x}Fe_xO_{3-\delta}$ perovskite oxide electrodes. *J Mater. Chem. A* **13**, 9708-9714 (2025).

36. Smith, D. W. An acidity scale for binary oxides. *J. Chem. Edu.* **64**, 480–481 (1987).

37. Guo, X. & Waser, R. Electrical properties of the grain boundaries of oxygen ion conductors: Acceptor-doped zirconia and ceria. *Prog. Mater. Sci.* **51**, 151–210 (2006).

38. Guo, X. & Waser, R. Space charge concept for acceptor-doped zirconia and ceria and experimental evidences. *Solid State Ion.* **173**, 63–67 (2004).

39. Toyama, S., Seki, T., Feng, B., Ikuhara, Y. & Shibata, N. Direct observation of space-charge-induced electric fields at oxide grain boundaries. *Nat. Commun.* **15**, 1–9 (2024).

40. Tschöpe, A., Kilassonia, S., Zapp, B. & Birringer, R. Grain-size-dependent thermopower of polycrystalline cerium oxide. *Solid State Ion.* **149**, 261–273 (2002).

41. Kim, S. & Maier, J. On the Conductivity Mechanism of Nanocrystalline Ceria. *J. Electrochem. Soc.* **149**, J73 (2002).

42. Chiang, Y. M., Lavik, E. B., Kosacki, I., Tuller, H. L. & Ying, J. Y. Nonstoichiometry and Electrical Conductivity of Nanocrystalline $CeO_{2-x}$. *J. Electroceram.* **1**, 7–14 (1997).

43. Defferriere, T. *et al.* Grain Boundary Space Charge Engineering of Solid Oxide Electrolytes: Model Thin Film Study. *arXiv:*2504.10684. (2025).

44. Nicollet, C. & Tuller, H. L. Perspective on the Relationship between the Acidity of Perovskite Oxides and Their Oxygen Surface Exchange Kinetics. *Chem. Mater.* **34**, 991–997 (2022).

## Acknowledgements

This work was supported by the National Research Foundation of Korea (NRF) grant funded by the Korea government (MSIT) (RS-2024-00341773) and by the Sagol Weizmann Institute of Science - MIT Bridge Program (2750231).

## Author information

### Corresponding Authors

*E-mail: hltuller@mit.edu

### Author Contributions

G.R.L. and H.L.T. conceived the original project. G.R.L., T.D., H.G.S., and H.L.T. designed the experimental protocol. G.R.L. conducted the overall sample preparation, the physicochemical property characterization, and electrochemical measurements. T.D. proposed the modeling of the electronic properties results. G.R.L. and T.D. analyzed the experimental and computational results. J.K. fabricated the target source for PLD and performed the deposition. H.L.T. supervised the project. G.R.L., T.D., Y.S.J., and H.L.T. wrote and reviewed the manuscript with input from all the authors. All authors discussed the results and contributed to revisions of the manuscript.

## Ethics declarations

### Competing interests

The authors declare no competing financial interests.

## Additional information

Supplementary Information is available for this paper at https://

# Supplementary Information

# Acidity-Mediated Metal Oxide Heterointerfaces: Roles of Substrates and Surface Modification

Gyu Rac Lee[1+], Thomas Defferriere[1+], Jinwook Kim[3], Han Gil Seo[4], Yeon Sik Jung[2], and Harry L. Tuller[1*]

*[1]Department of Materials Science and Engineering, Massachusetts Institute of Science and Technology, Cambridge, MA 02139, USA*
*[2]Department of Materials Science and Engineering, Korea Advanced Institute of Science and Technology, 291 Daehak-ro, Yuseong-gu, Daejeon 34141, Republic of Korea*
*[3]Department of Materials Science and Engineering, Northwestern University, Evanston, IL 60208, USA*
*[4]Department of Materials Science and Engineering, Dankook University, 119 Dandae-ro, Dongnam-gu, Cheonan-si, Chungnam 31116, Republic of Korea*

*[+]These authors contributed equally: Gyu Rac Lee, Thomas Defferriere*

## Table of Contents

**Supplementary Note 1. Fabrication of $PCO_{NA}$ by solvent-assisted nanotransfer printing combined with pulsed laser deposition**

$Pr_{0.2}Ce_{0.8}O_{2-\delta}$ (PCO20) was selected as a model MIEC oxide due to its high chemical stability[1], free of elements that tend to segregate to the surface, for example Sr in LSCF and STF[2–4], and the dominance of electronic over ionic conductivity at high $pO_2$ (e.g. in this study in air)[5,6]. In conventional solvent-assisted nanotransfer printing (S-nTP), electron beam evaporation is typically used to deposit the target materials, however, this is not suitable for depositing complex oxide materials normally used as MIEC electrodes such as PCO20. Electron beam irradiation, when applied to metal oxides, leads to the reduction and dissociation of the target source into metal and metal suboxide species, each with different volatilities, thereby hindering the reproduction of the stoichiometry of the target source in the deposited film[7,8]. Pulsed laser deposition (PLD) was therefore selected as an alternative to electron beam evaporation, given its ability to reproduce films with nearly the same stoichiometry as the target source[9]. It is worth noting that this strategy is not limited to PCO used in this study but is extendable to other materials that can be deposited via PLD. We believe that our strategy exhibits wide-ranging applicability and potential for elucidating the precise role of acidity-mediated local space charge across various materials and applications.

**Supplementary Note 2. Modeling of heterointerface acidity-mediated space charge effects on electronic properties of $PCO_{NA}$**

Zhang et al. previously showed that decreasing the grain size of a bulk PCO20 sample by lowering the sintering temperature led to a corresponding decrease in conductivity and in $E_a$ from 1 eV down to 0.65 eV. They attributed this to increased strain, that lowers the reduction enthalpies, increases the $[Pr^{3+}]$ concentration, and therefore reduces the conductivity by decreasing the availability of small polaron hopping sites[10]. As we show in the Arrhenius plot of **Supplementary Fig. 8**, $PCO_{NA}$ prepared on an $Al_2O_3$ substrate and annealed at 400 °C instead of 700 °C, exhibited similar $E_a$ and magnitude of conductivity comparable to that of bulk PCO20, despite the sample being composed of nanoscale grains that are presumably smaller due to the lower annealing temperatures.

To rationalize our observations, we note that the activation energy of conductivity ($E_a$) in bulk PCO is reportedly made up of two terms, given that the electron concentration and mobility are both thermally activated. Following previous derivations[10], $[Pr_{Ce}^{x}] \sim [Pr_{Ce,tot}]$ is assumed in the high $pO_2$ regime where overall charge balance is assumed to be $[Pr_{Ce}'] \sim 2[V_{O}^{\cdot\cdot}]$. Then the small polaron electronic conductivity associated with $Pr^{3+}$ can be written as:

$$\sigma_{Pr^{3+}} = 2^{\frac{1}{3}}[Pr_{Ce,tot}]^{\frac{2}{3}}[O_O^x]^{\frac{1}{3}}K_{r,Pr}^{0}{}^{\frac{1}{3}}\mu_{Pr,0}\exp\left(-\frac{\frac{\Delta H_{r,Pr}}{3}+E_{m,Pr}}{k.T}\right)P_{O_2}^{-\frac{1}{6}} \quad (1)$$

where $\Delta H_{r,Pr}$ is the reduction enthalpy of Pr and $E_{m,Pr}$ is the small polaron migration energy. Based on previous literature values, the expected bulk activation energy at high $pO_2$ and low temperatures is therefore expected to be equal to $\frac{\Delta H_{r,Pr}}{3} + E_{m,Pr} \sim 1$ eV, consistent with our results for $PCO_{NA}$ fabricated on $Al_2O_3$ substrate annealed at 400 °C (**Supplementary Fig. 8**), where little impact from GB space charge effects was anticipated.

On the other hand, the carrier concentration is also expected to vary spatially according to the magnitude and polarity of the space charge potentials at both heterointerfaces and GBs. Since small polarons are mobile, they can readily redistribute within the space charge region in response to changes in space charge potential, similarly to the oxygen vacancies, a situation that can be described by the Gouy-Chapman approximation[11], where a flat spatial profile of $[Pr_{tot}]$ is assumed. To a first-order approximation, the variation in $Pr_{Ce}'$ adjacent to the interface in response to the space charge potential profile can be defined as[12] :

$$\frac{[Pr'_{Ce}](x)}{[Pr'_{Ce}]_{bulk}} = exp\left(-\frac{ze}{kT}\Delta\varphi(x)\right) \quad (2)$$

where the potential distribution $\Delta\varphi(x)$, can be obtained by solving the Poisson equation to account for the total electroneutrality between the interface and the space charge layer[12]. An analytical solution is generally possible by making the depletion approximation, whereby the concentration of one of the carriers is negligible compared to the other (generally true when the two major carriers are of opposite sign, causing the depletion of one in the space charge region and the accumulation of the other).

In general, we can then define conductance parallel and perpendicular to the interface as:

$$\sigma_{//,sc} = \frac{1}{L_D}\int_0^{2L_D}[Pr^{3+}](x)\left(1-\frac{[Pr^{3+}](x)}{[Pr_{tot}]}\right)q\mu_{Pr}\,dx \quad (3)$$

$$\rho_{\perp,sc} = \frac{1}{L_D}\int_0^{2L_D}\frac{1}{[Pr^{3+}](x)\left(1-\frac{[Pr^{3+}](x)}{[Pr_{tot}]}\right)q\mu_{Pr}}dx \quad (4)$$

where $L_D = \left(\frac{kT\varepsilon}{2e^2[Pr'_{Ce}]_{bulk}}\right)^{0.5}$ is the Debye length and $\left(1-\frac{[Pr^{3+}](x)}{[Pr_{tot}]}\right) = \frac{[Pr^{4+}](x)}{[Pr_{tot}]}$ reflects the fact that in PCO the mobility of localized electrons sitting on $Pr^{3+}$ sites is only possible if adjacent available Pr sites are "empty" ($Pr^{4+}$ state). The measured $E_a$ in the space charge regions is thus expected to have multiple contributions, which include the sum of one-third of the reduction enthalpy of Pr in PCO, the small polaron hopping energy, and a contribution from the space charge potential. Directly solving these equations can become challenging especially in the scenario that $\Delta\phi > 0$ where we can not simplify the site occupancy term $\left(1-\frac{[Pr^{3+}](x)}{[Pr_{tot}]}\right)$. Moreover, one of the major assumptions in deriving the Gouy-Chapman model no longer holds, i.e., the assumption of a dilute limit condition. Nevertheless, for the sake of our study, we can visually inspect equations (3) and (4) to predict the expected behavior.

When $\Delta\phi < 0$, a situation described at the substrate/surface heterointerface of our nanowire in contact with an acidic oxide, then $\frac{[Pr'_{Ce}](x)}{[Pr'_{Ce}]_{bulk}} < 1$ and $[Pr'_{Ce}](x)$ becomes depleted in the vicinity of the interface. Conduction along $\sigma_{//,sc}$ would systematically be bypassed by the parallel contribution of $\sigma_{bulk}$ (**Supplementary Fig. 6a**).

When $\Delta\phi > 0$, a situation described at the substrate/surface heterointerface when our nanowire is in contact with a basic oxide or at the grain boundaries, then an accumulation of $Pr^{3+}$ in the space charge region is expected to occur (**Supplementary Fig. 6b**). In the accumulated space charge region, where $[Pr^{3+}](x) \sim 50\%[Pr_{tot}]$, then a local maximum in conductivity is expected, where $[Pr^{3+}](x)\left(1-\frac{[Pr^{3+}](x)}{[Pr_{tot}]}\right)$, is maximized. Above this point, closer to the interface, the effective mobility of small polaron is expected to decrease with further increases in $[Pr^{3+}](x)$, owing to the decrease in site availability (i.e. $[Pr^{3+}](x)\left(1-\frac{[Pr^{3+}](x)}{[Pr_{tot}]}\right)$ . The implication is that an anisotropic conductivity profile may arise, with a high resistivity region developing close to the interface wherever $[Pr^{3+}](x) > 50\%[Pr_{tot}]$, contributing to a series resistance for perpendicular transport across the interface, while a conductivity maxima where $[Pr^{3+}](x) \sim 50\%[Pr_{tot}]$, occurring a small distance away from the interface for parallel transport. The conductivity profile will follow the trends displayed in **Supplementary Fig. 7**:

The above rationale explains how electron accumulation at a heterointerface can enable enhanced conductivity parallel to the interface (for MgO and $Li_2O$), while causing higher resistance for bulk transport across grain boundaries. It is also entirely consistent with the observations made previously by Zhang et al. in bulk nanocrystalline ceramic of PCO, when we consider that ceramic samples sintered at higher temperatures typically exhibit smaller grain boundary space charge potentials[10]. Nanocrystalline samples sintered at lower temepratures are therefore expected to have higher space charge potentials and therefore stronger $Pr^{3+}$ accumulation creating a more significant resistive blockage to perpendicular transport across the space charge region, even though the apparent activation energy is reduced.

Moreover, we recently showed that the GB space charge potentials in acceptor-doped $CeO_2$ thin films could be modulated through the in-diffusion of elements from the substrates at intermediate temperature (700-900 °C)[13]. We showed that depending on the element diffusing up to the GB (Al/Mg), increases or decreases in space charge potentials would be observed. This was enabled by the fact that cation diffusion along GB cores is accelerated relative to the bulk at these intermediate temperatures and the fact that $Al^{3+}$ ions are believed to sit on interstitial sites in GBs in ceria-based ceramics, inducing a net positive charge of 3+ ($Al_i^{\cdot\cdot\cdot}$) and therefore an increase in space charge potential[14–16], while $Mg^{2+}$ sits substitutionally, leading to a net negative charge of 2- ($Mg_{Ce}''$)[13]. The up-diffusion of $Mg^{2+}$ would be expected to decrease the space charge potential at the grain boundaries, while leading to a depletion of electrons at the substrate/surface interface. This grain boundary in-diffusion process would also explain why transport across grain boundaries will be more resistive than the bulk while possessing a lower activation energy, as the sign of the space charge potential would actually act to reduce the total activation of conductivity composed of one-third of the reduction enthalpy and bulk mobility (note that in equation(1) and equation (2), the exponential terms have different signs). In the case of **Supplementary Fig. 8**, where the $PCO_{NA}$ on $Al_2O_3$ is annealed at 700°C, $Al^{3+}$ at the grain boundary is expected to cause a highly increased space charge potential and therefore a higher series resistance ($\rho_{\perp}$) induced by the depressed small polaron mobility, contributing to a lower conductance. On the other hand at 400°C, the space charge potential at the grain boundaries is expected to be small and therefore the bulk conductivity of the film would align closer with bulk theoretical expectations. In **Supplementary Fig. 7**, we provide a 3D visualization of the anisotropic charge transport that would occur at surface and grain boundaries due to a $\Delta\phi > 0$ and compare it to the situation of electron depletion at the substrate interface (i.e. $\Delta\phi < 0$).

**Supplementary Note 3. Influence of insulating substrate acidity on 3 mol% Gd-doped ceria (GDC3) nanowire arrays ionic properties.**

In the case of GDC3, where ionic conduction is the dominant transport mechanism, which occurs through oxygen vacancy hopping, the conduction behavior differs from that of $PCO_{NA}$. Specifically, when a relatively basic MgO was used as a substrate, oxygen vacancy depletion was induced, leading to a total decrease in conductivity. On the other hand, the $Al_2O_3$ (relatively acidic) substrate promoted vacancy accumulation, resulting in a total increase in conductivity. In contrast to PCO, where bulk transport is composed of contributions from thermal carrier generation and their migration, in GDC3, the bulk conductivity is composed only of the migration energy of the oxygen vacancies (- 0.7 eV) as the carrier concentration is fixed by the aliovalent $Gd^{3+}$ dopant, while for polycrystalline samples, activation energies > 1 eV are typically reported, associated with the additional grain boundary series resistance caused by the space charge barriers with positive potentials barriers[11,13].

The ionic properties of GDC3 nanowire arrays on two different substrates were characterized through Arrhenius plots in **Supplementary Fig. 10**. In this case, both samples were annealed at 400 °C, so the GB space charge potential driven by cation in-diffusion from substrate can be ignored. GDC3 nanowire arrays on the MgO substrate exhibited similar $E_a$ (1.23 eV) comparable to that of polycrystalline GDC3 thin film. However, GDC3 nanowire arrays on the $Al_2O_3$ substrate exhibited a lower $E_a$ of 1.01 eV. Moreover, GDC3 nanowire arrays on the $Al_2O_3$ showed higher in-plane conductivity than those on the MgO substrate.

These results are attributed to the conduction pathways determined by the acidity-mediated space charge potential at the heterointerface. In the case of MgO, a resistive heterointerface is formed, causing conduction to occur through the bulk ($R_1$). On the other hand, in the case of $Al_2O_3$, oxygen vacancy accumulation facilitates conduction through the substrate heterointerface ($R_2$), which is opposite to the tendency observed in $PCO_{NA}$. Furthermore, for $Al_2O_3$, the electric fields generated by the space charge at the heterointerface (due to a negative space charge potential) are opposite directions to the positively charged space charge at the GB, partially compensating each other and leading to a reduced $E_a$ at the grain boundary heterointerface triple junction.

**Supplementary Note 4. Comparison of polycrystalline and epitaxial PCO20 films to elucidate grain boundary in-diffusion effects**

Polycrystalline and epitaxial PCO20 films were fabricated using PLD with a PCO20 target on $Al_2O_3$ substrate by modulating the temperature and deposition rate. For the polycrystalline PCO20 film, deposition was conducted at room temperature with a laser frequency of 10 Hz. In contrast, the epitaxial film was deposited at 700 °C with a reduced frequency of 2 Hz, while all other deposition conditions were maintained constant.

Evidence for the role of GB space charge potentials in impacting the effective conductivity of polycrystalline films can be obtained by comparing the AC impedance analysis of polycrystalline and epitaxial PCO20 films fabricated by PLD. We note that similarly grown polycrystalline and epitaxial films of PCO20 exhibited a trend (decrease in conductivity and activation energy) when moving from the epitaxial film to polycrystalline film, though the activation energies were generally lower than for the PCO20 bulk reference, likely due to lattice strain. In polycrystalline films, numerous GBs are present, allowing Al to diffuse into them during annealing. In contrast, in epitaxial films without GBs, such space charge modulation is prevented even under high-temperature annealing conditions (**Supplementary Fig. 12d**). As verified by XRD analysis (**Supplementary Fig. 12a and 12b**), the epitaxial films predominantly exhibit (111) and its family of planes, while polycrystalline films display a mix of planes, including (111) and (200). In addition, the shifted positions of peaks appearing in both polycrystalline and epitaxial films compared to $PCO_{NA}$ imply tensile strain, which is reported to decrease the migration energy and improve diffusivity of oxygen vacancy hopping[10,17]. As expected, the total conductivity of the epitaxial film is higher than that of the polycrystalline film due to increased blocking at the polycrystalline film's GBs. Furthermore, the Arrhenius plots reveal that the $E_a$ of the polycrystalline film is lower than that of epitaxial films associated with the space charge regions surrounding the GBs, consistent with the $PCO_{NA}$ results (**Supplementary Fig. 12c**).

**Supplementary Table 1.** Measurement of Pr chemical composition of $PCO_{NA}$ fabricated using a PCO 20 at % target, through inductively coupled plasma mass spectrometer (ICP-MS). The reported loading amount of Pr in the $PCO_{NA}$ represents the mean ± standard deviation (*n* = 3)

| **Sample** | **ICP-MS (at %)** |
|---|---|
| $PCO_{NA}$<br>(Fabricated using PCO 20 at % target) | 19.15 ± 0.1 |

**Supplementary Table 2.** Absolute conductivity values for the binary oxides ($Li_2O$ and $SiO_2$) at 600 °C used in this work.

| | **Conductivity (S/cm) at 600 °C** |
|---|---|
| $Li_2O$ (for basic)[18] | - $10^{-5}$ |
| $SiO_2$ (For acidic)[19] | - $10^{-12}$ |

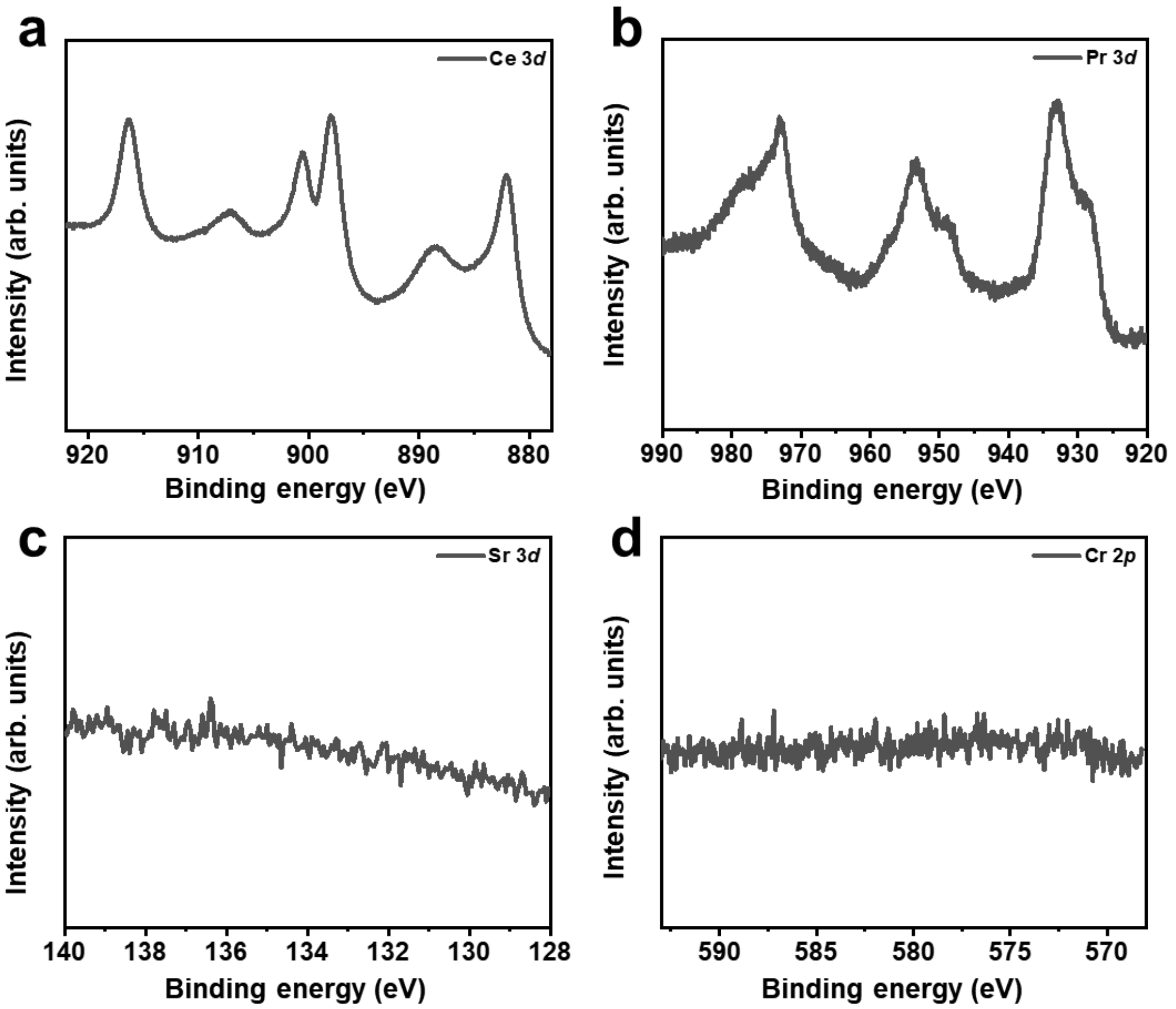

**Supplementary Fig. 1** X-ray photoelectron spectroscopy (XPS) spectra of pristine $PCO_{NA}$ with the different binding energy regions of Ce 3*d*, Pr 3*d*, Sr 3*d*, and Cr 2*p* spectra, respectively. XPS data exhibits no presence of Sr or Cr elements at the surface which could cause surface degradation in $PCO_{NA}$ sample.

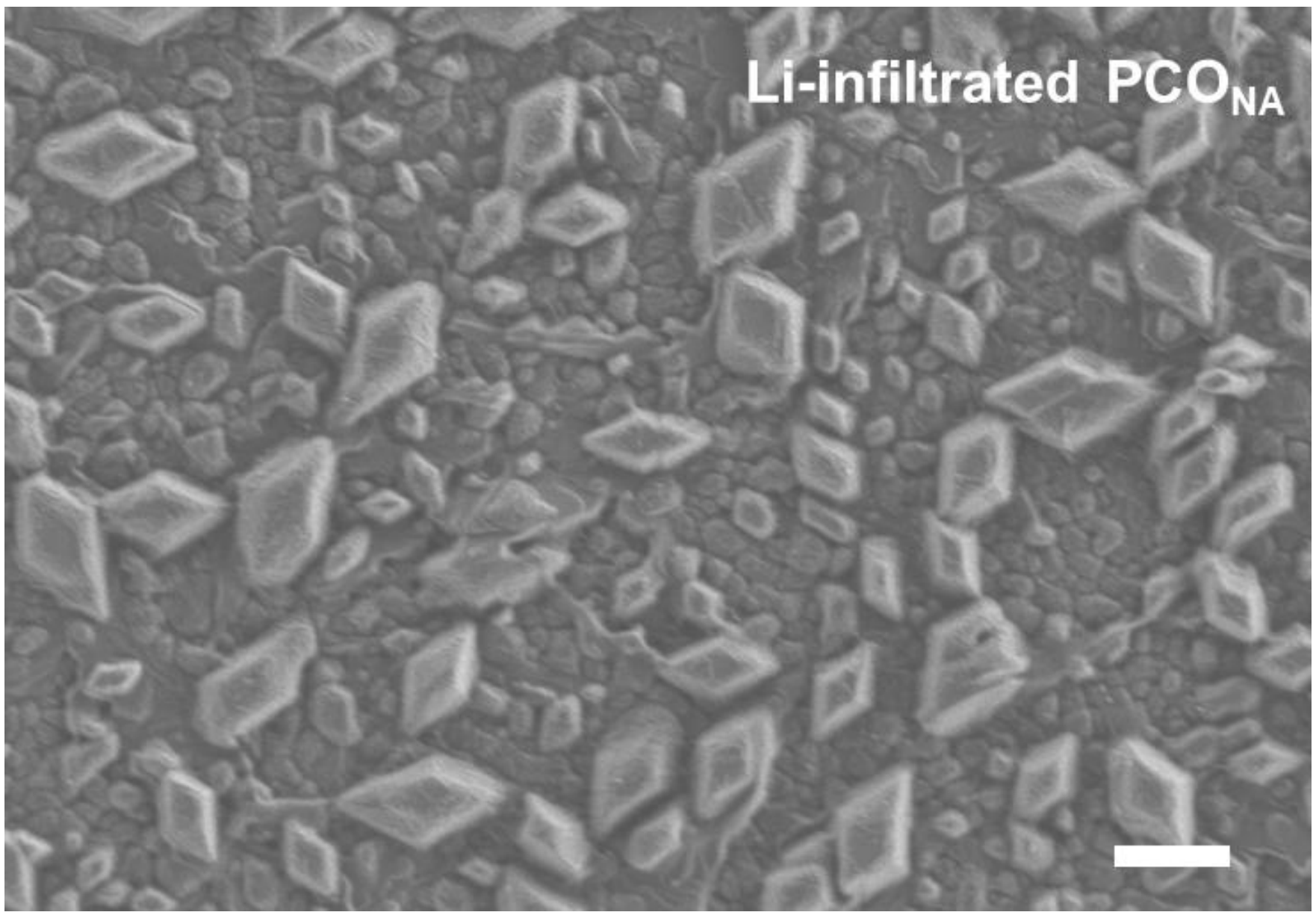

**Supplementary Fig. 2** SEM image of Li-infiltrated $PCO_{NA}$ fabricated on $Al_2O_3$. The $PCO_{NA}$ is observed to be fully covered by $Li_2O$ given infiltration with a concentrated solution. (scale bar, 2 $\mu$m).

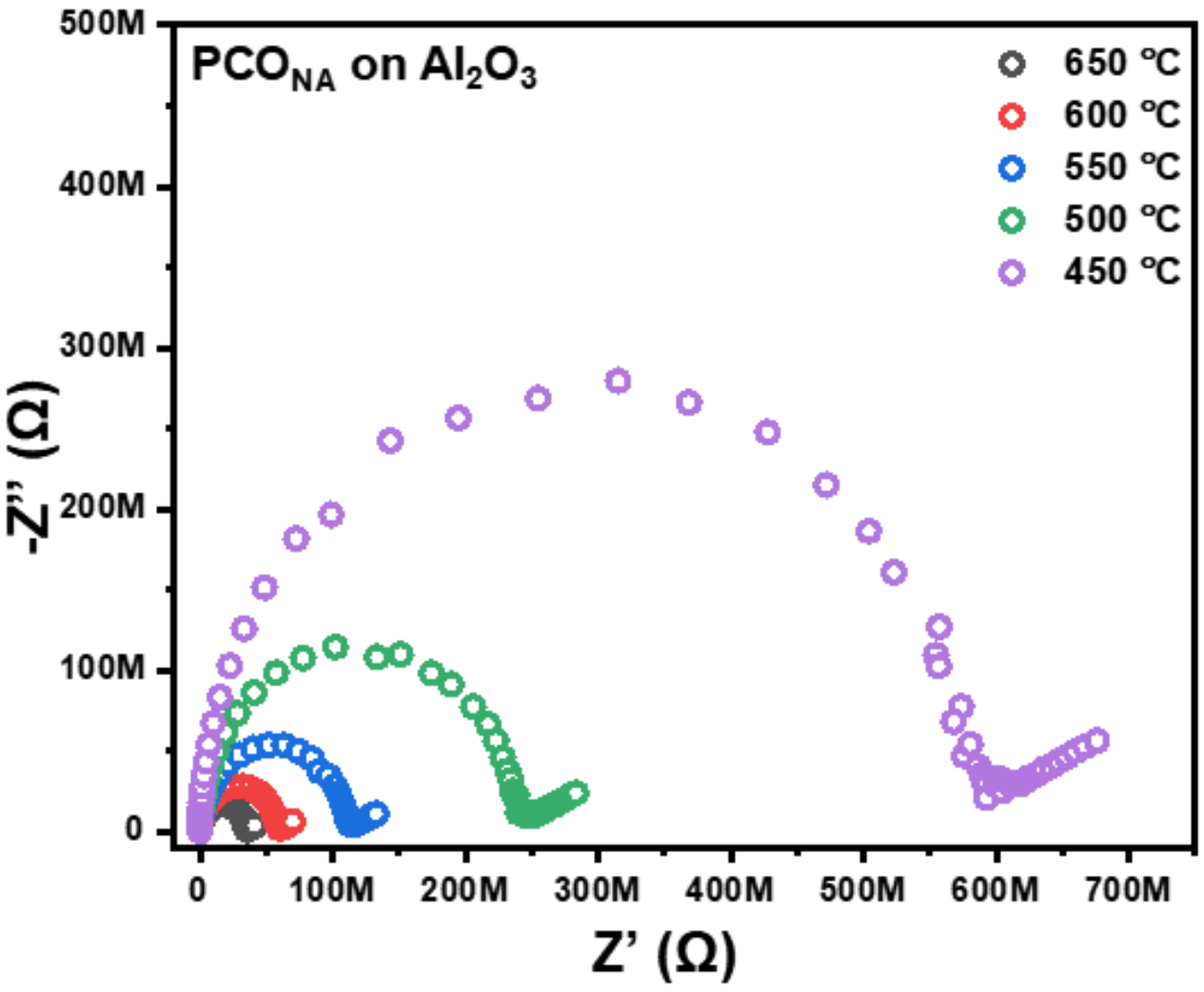

**Supplementary Fig. 3** Impedance spectra measured from 450 °C to 650 °C on uninfiltrated PCO$_{NA}$ fabricated on $Al_2O_3$ insulating substrate.

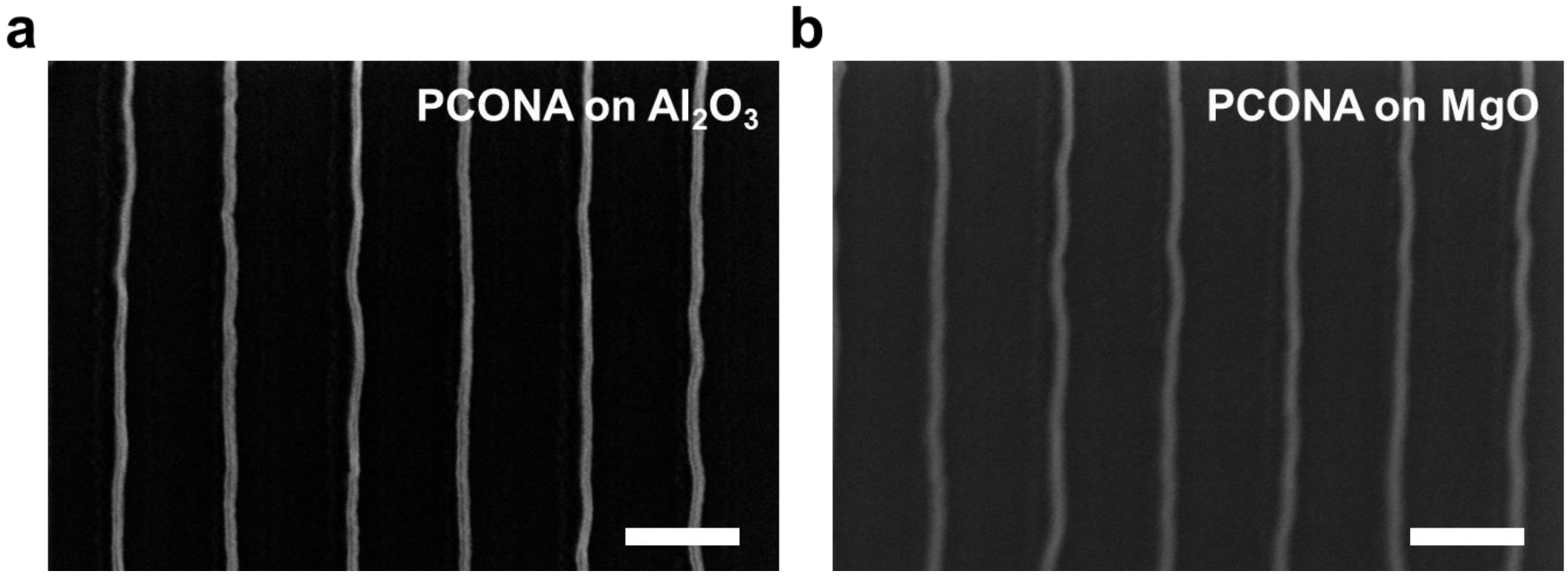

**Supplementary Fig. 4** Scanning electron microscopy (SEM) images of $PCO_{NA}$ fabricated on **a** $Al_2O_3$ and **b** MgO substrates, respectively (scale bar, 200 nm).

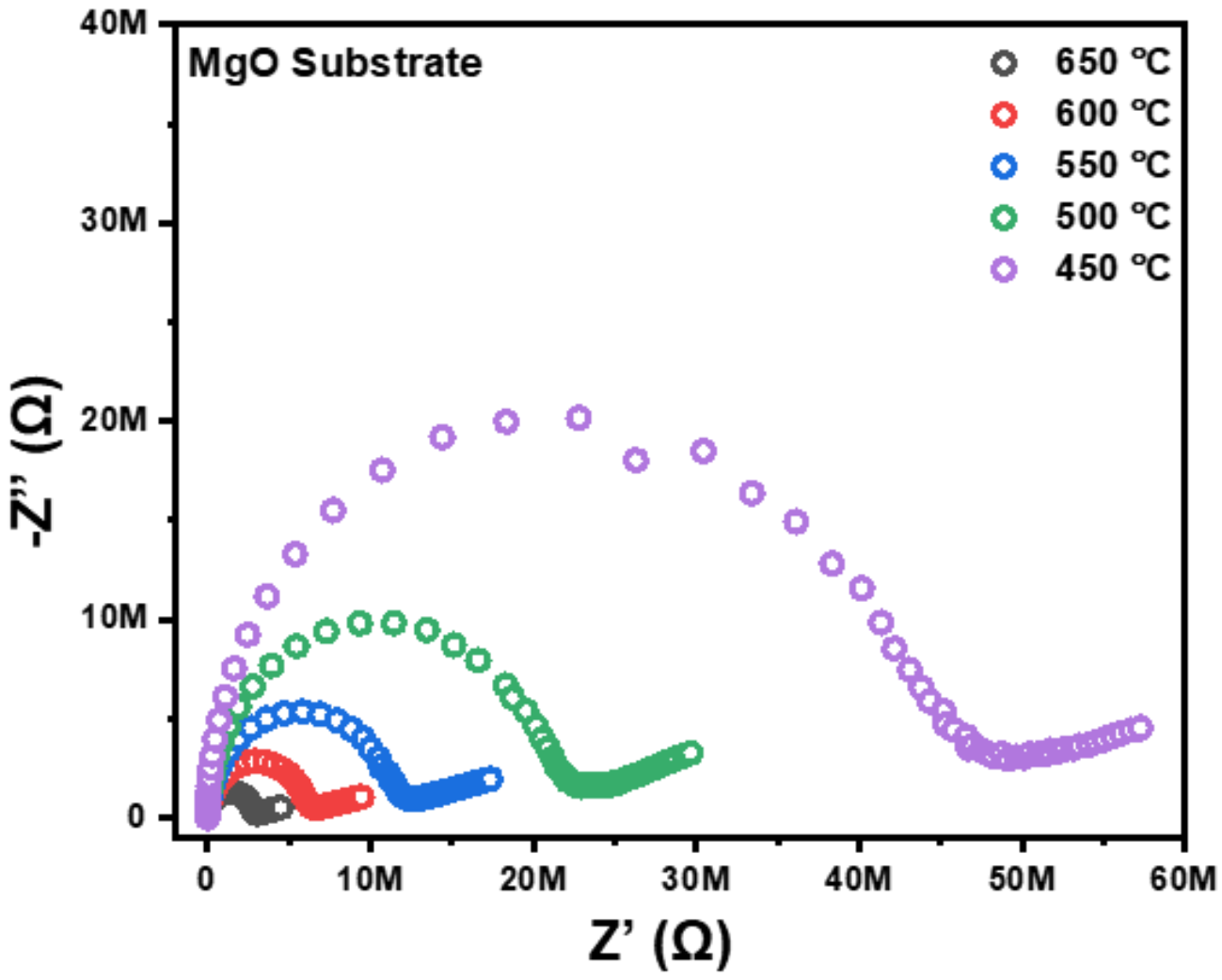

**Supplementary Fig. 5** Impedance spectra measured from 450 °C to 650 °C on uninfiltrated $PCO_{NA}$ fabricated on MgO insulating substrate.

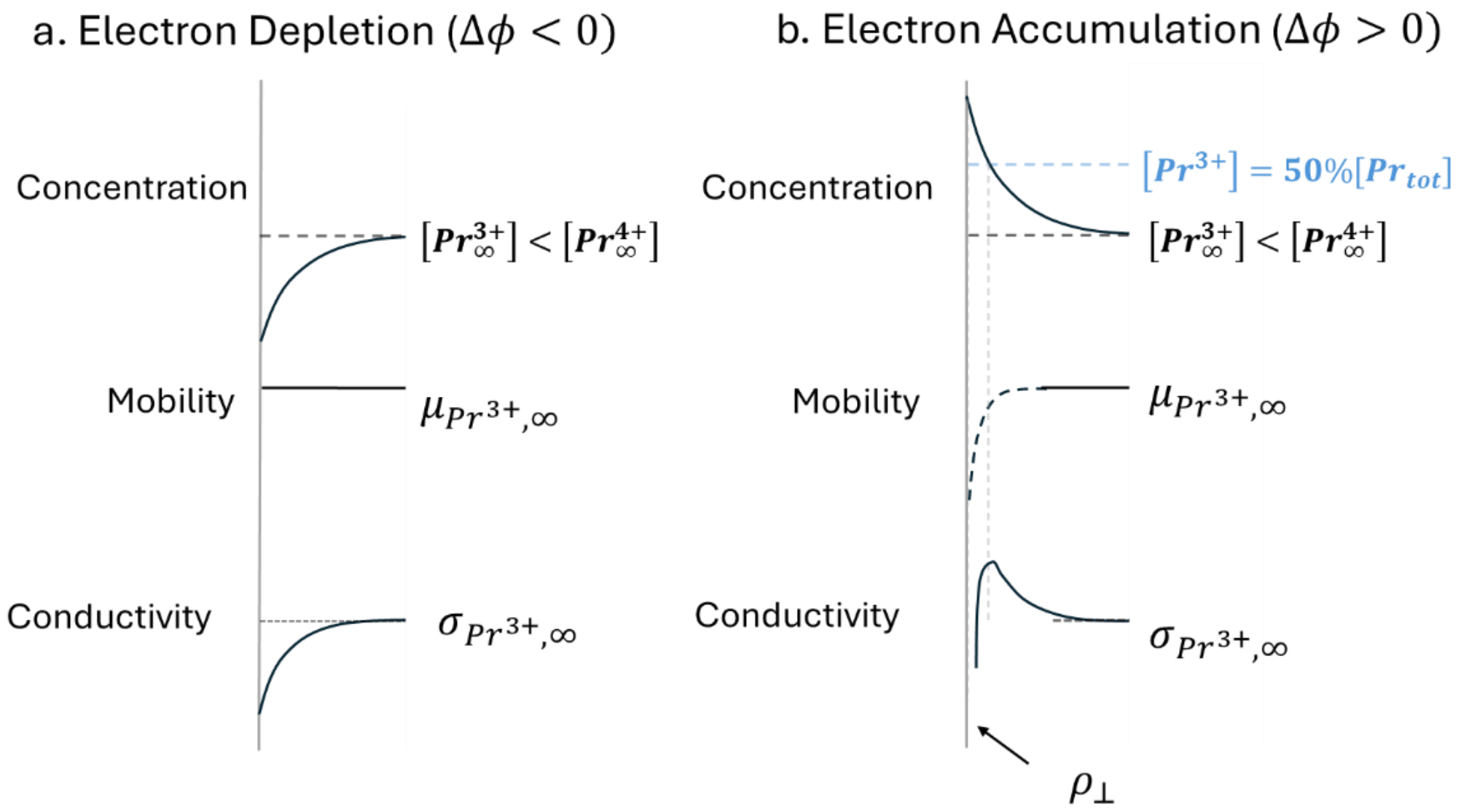

**Supplementary Fig. 6** Illustration of space charge model for the case of a. positive and b. negative space charge potential and its impact on the spatial distribution in space charge region of $Pr^{3+}$ concentration ( $[Pr^{3+}]$ ), effective mobility $\mu_{Pr,effective} \sim [Pr^{3+}](x)\big(1 - [Pr^{3+}](x)\big)\mu_{Pr}$ and expected conductivity ($\sigma$).

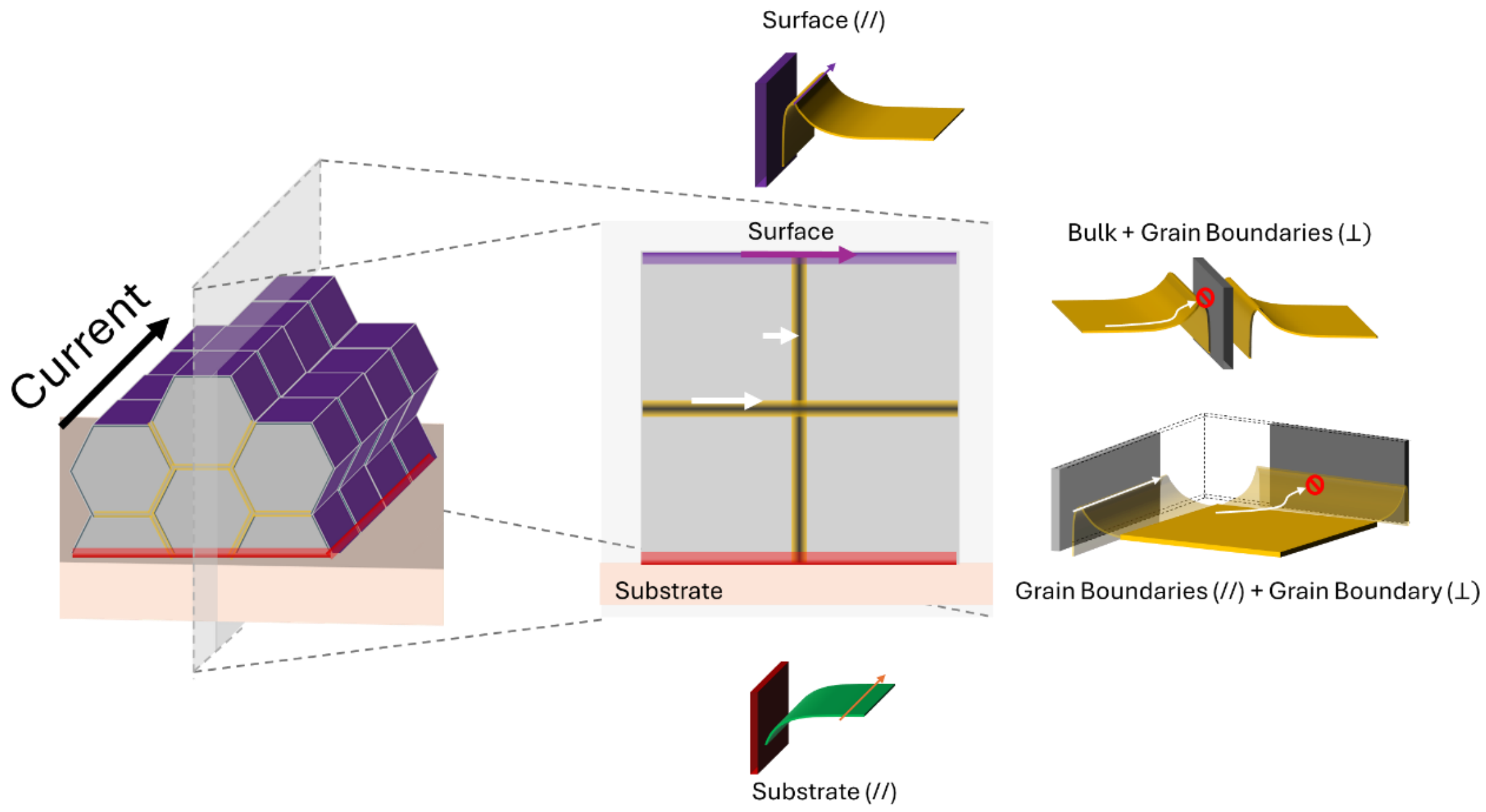

**Supplementary Fig. 7** 3D visualization of nanowire composed of multiple nanoscale grains. (left image) cross section of a nanowire – (center image) along nanowire. Purple shaded and gray areas surrounded by yellow shaded areas, represent the surface in contact with a base and grain boundaries where $\Delta\phi > 0$, causing an anisotropic profile of $[Pr^{3+}](x)$ enabling parallel conduction to the surface (//), while causing resistive conduction across the grain boundaries (⊥). Red shaded areas at the bottom in contact with substrate represent case of nanowire in contact with acidic susbtrate (i.e. $Al_2O_3$), expected to cause $\Delta\phi < 0$ and therefore electron depletion.

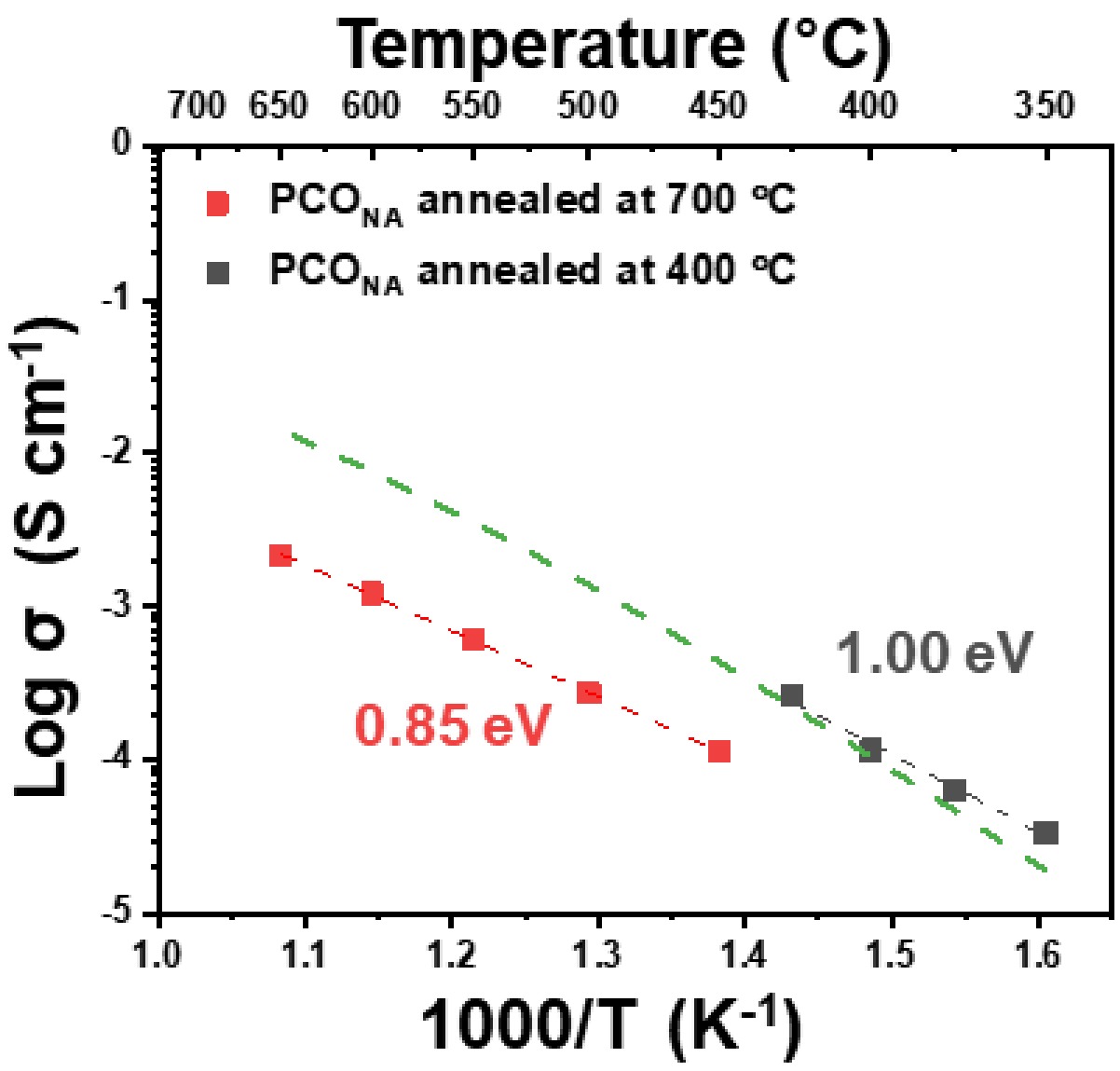

**Supplementary Fig. 8** Arrhenius plots of in-plane conductivity of $PCO_{NA}$ fabricated on $Al_2O_3$ substrates annealed at 400 °C and 700 °C, respectively. The green dashed line indicates the conductivity behavior of bulk PCO20 calculated with assistance of defect chemical modelling.

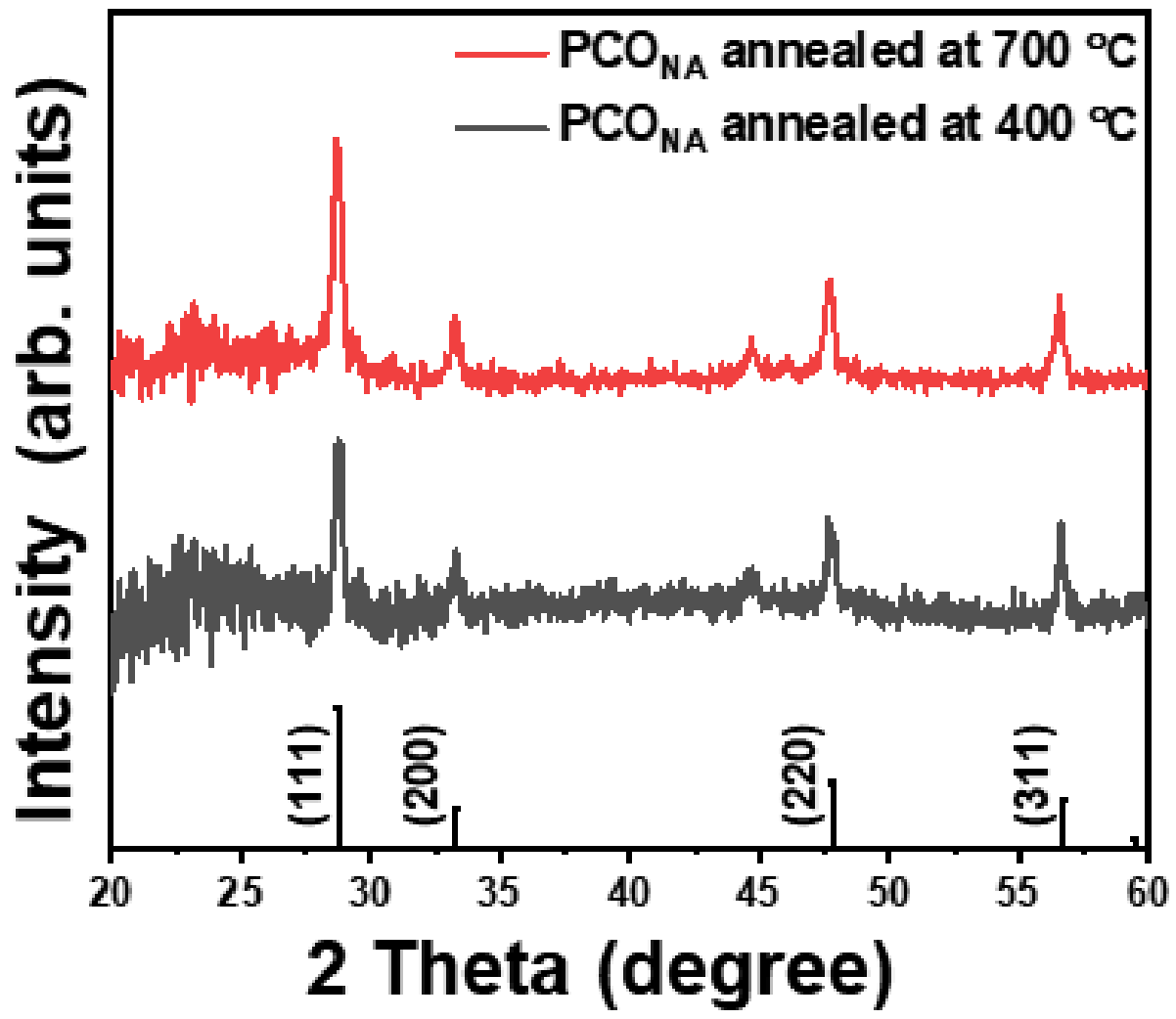

**Supplementary Fig. 9** XRD spectra of $PCO_{NA}$ fabricated on $Al_2O_3$ substrate annealed at 400 °C and 700 °C, respectively. Black vertical lines at bottom of figure indicate the dominant facets of the cerium oxide crystal structure.

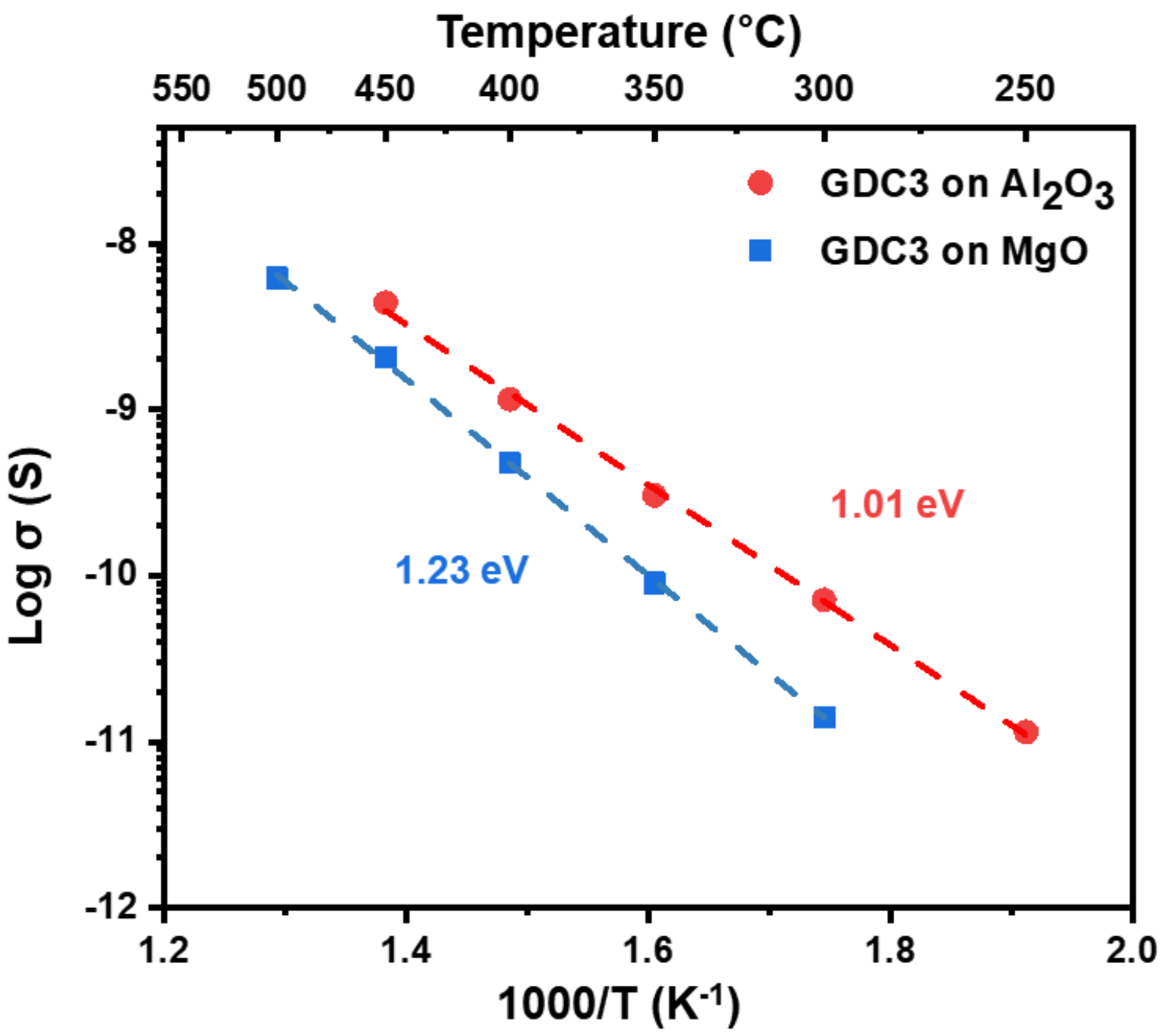

**Supplementary Fig. 10** Arrhenius plots of in-plane conductivity of 3 at% Gd-doped ceria (GDC3) nanowire arrays fabricated on $Al_2O_3$ and MgO substrates annealed at 400 °C along with their respective activation energies.

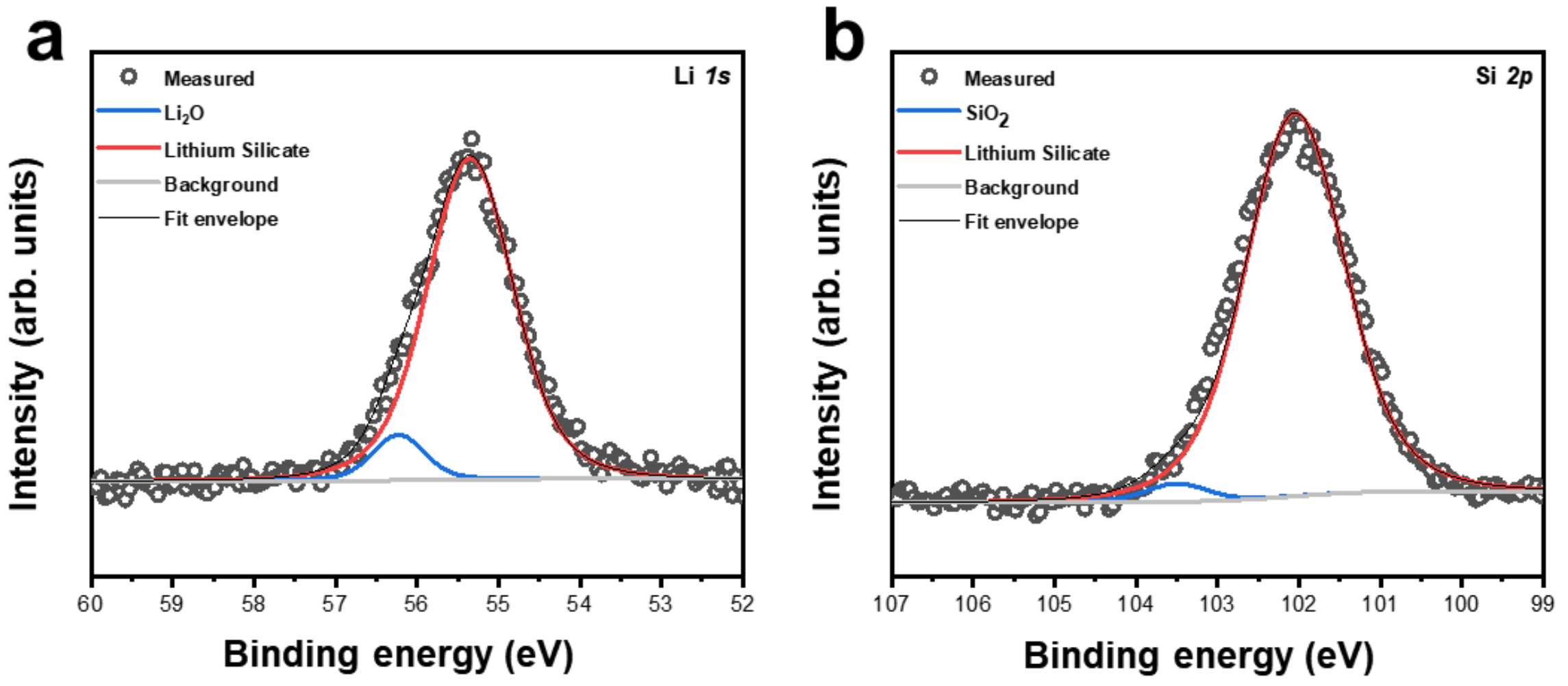

**Supplementary Fig. 11** XPS spectra of $PCO_{NA}$ after sequential infiltration with Li-species following Si-species, showing the different binding energy regions of Li *1s* and Si *2p* spectra, respectively.

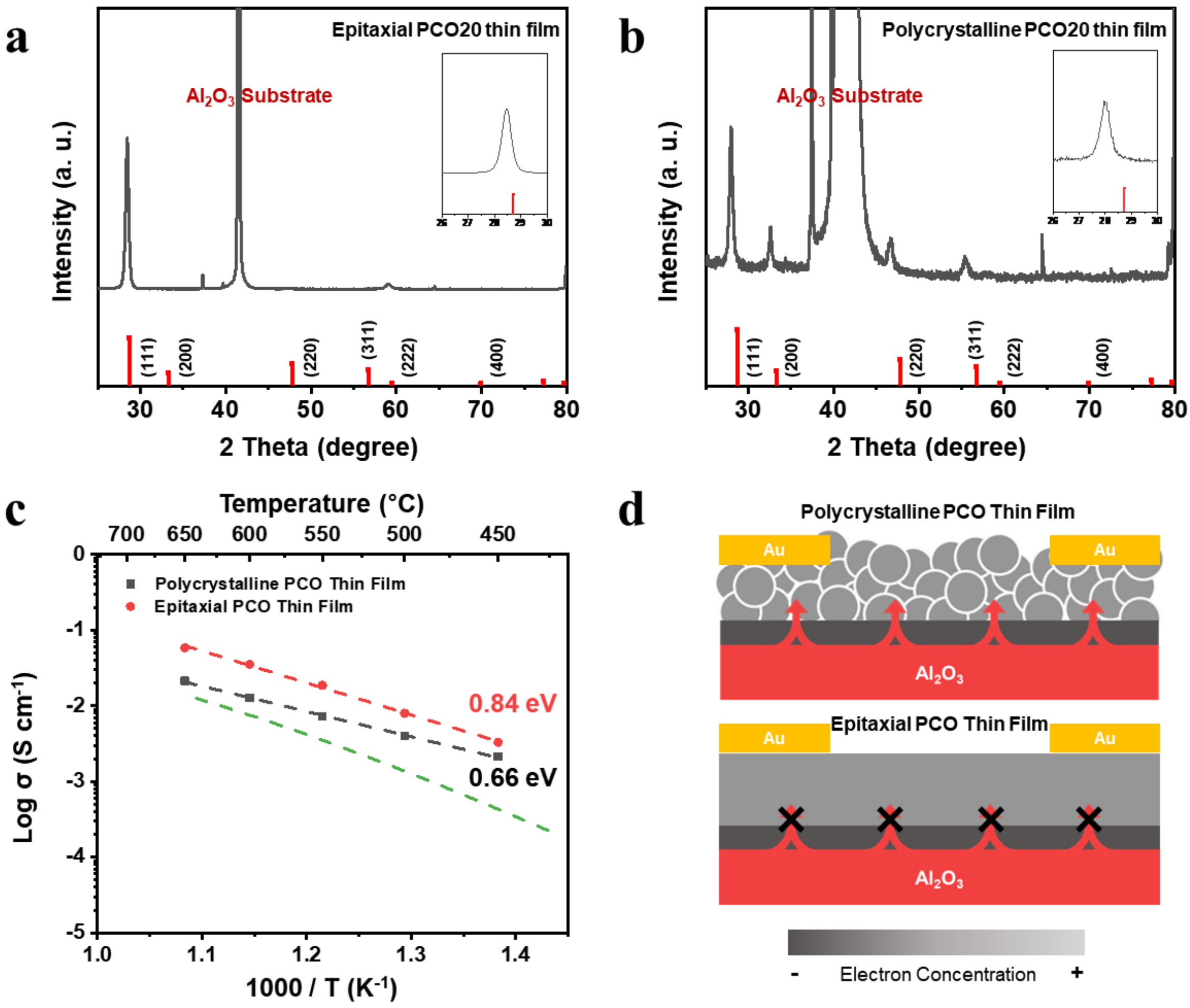

**Supplementary Fig. 12** XRD spectra of **a** epitaxial and **b** polycrystalline PCO20 thin films deposited on $Al_2O_3$ insulating substrates via pulsed laser deposition (PLD). The inset images show enlarged views of the spectra from 26° to 30°, reavealing a peak shift due to tensile strain. Red vertical lines at bottom of figures indicate dominant facets of cerium oxide crystal structure. **c** Arrhenius plots of in-plane conductivity of epitaxial and polycrystalline PCO20 thin films, along with their respective activation energies. The green dashed line indicates the conductivity behavior of a bulk PCO20 film calculated with the aid of defect chemical modelling. **d** Comparison between polycrystalline and epitaxial PCO20 thin films. Substrate heterointerface shown in black corresponds to electron depletion due to relative acidity of $Al_2O_3$. Diffusion of Al ions only occurs in the polycrystalline PCO20 thin film, leading to GB electron accumulation, depicted in white.

## References

1. Takada, R. & Yao, H. Praseodymium-Doped Cerium Oxide (PrxCe1-xO2−δ) Nanoparticles with High Water Dispersibility: The Nature of Pr-Related Optical Transitions Studied by MCD Spectroscopy. *J. Phys. Chem. C* **129**, 5461–5471 (2025).

2. Koo, B. *et al.* Sr Segregation in Perovskite Oxides: Why It Happens and How It Exists. *Joule* **2**, 1476–1499 (2018).

3. Kim, H., Seo, H. G., Ahn, S., Tuller, H. L. & Jung, W. C. Unveiling critical role of metal oxide infiltration in controlling the surface oxygen exchange activity and polarization of $SrTi_{1-x}Fe_xO_{3-\delta}$ perovskite oxide electrodes. *J. Mater. Chem. A* **13**, 9708-9714 (2025).

4. Jung, W. & Tuller, H. L. Investigation of surface Sr segregation in model thin film solid oxide fuel cell perovskite electrodes. *Energy Environ. Sci.* **5**, 5370–5378 (2012).

5. Seo, H. G. *et al.* High-Performance and Durable Fuel Cells using Co/Sr-Free Fluorite-Based Mixed Conducting $(Pr,Ce)O_{2-\delta}$ Cathode. *Adv. Energy Mater.* **12**, 2202101 (2022).

6. Bishop, S. R., Stefanik, T. S. & Tuller, H. L. Defects and transport in $Pr_xCe_{1-x}O_{2-\delta}$: Composition trends. *J. Mater. Res.* **27**, 2009–2016 (2012).

7. Han, H. J. *et al.* Unconventional grain growth suppression in oxygen-rich metal oxide nanoribbons. *Sci. Adv.* **7**, eabh2012 (2021).

8. Shibata, N. *et al.* Direct imaging of reconstructed atoms on $TiO_2$ (110) surfaces. *Science* **322**, 570–573 (2008).

9. Schou, J. Physical aspects of the pulsed laser deposition technique: The stoichiometric transfer of material from target to film. *Appl. Surf. Sci.* **255**, 5191–5198 (2009).

10. Zhang, S., Fang, Z., Chi, M. & Perry, N. H. Facile Interfacial Reduction Suppresses Redox Chemical Expansion and Promotes the Polaronic to Ionic Transition in Mixed Conducting $(Pr,Ce)O_{2-\delta}$ Nanoparticles. *ACS Appl. Mater. Interfaces* **17**, 880-898 (2024).

11. Guo, X. & Waser, R. Electrical properties of the grain boundaries of oxygen ion conductors: Acceptor-doped zirconia and ceria. *Prog. Mater. Sci.* **51**, 151–210 (2006).

12. Tong, X., Mebane, D. S. & De Souza, R. A. Analyzing the Grain Boundary Resistance of Oxide Ion Conducting Electrolytes: Poisson-Cahn vs Poisson-Boltzmann theories. *J. Am. Ceram. Soc.* **103**, 5–22 (2020).

13. Defferriere, T. *et al.* Grain Boundary Space Charge Engineering of Solid Oxide Electrolytes: Model Thin Film Study. *arXiv:*2504.10684. (2025).

14. Xu, X. *et al.* Local Multimodal Electro-Chemical-Structural Characterization of Solid-Electrolyte Grain Boundaries. *Adv. Energy Mater.* **11**, 2003309 (2021).

15. Minervini, L., Zacate, M. O. & Grimes, R. W. Defect cluster formation in $M_2O_3$-doped $CeO_2$. *Solid State Ion.* **116**, 339–349 (1999).

16. Xu, X. *et al.* Variability and origins of grain boundary electric potential detected by electron holography and atom-probe tomography. *Nat. Mater.* **19**, 887–893 (2020).

17. Harrington, G. F. *et al.* Tailoring Nonstoichiometry and Mixed Ionic Electronic Conductivity in $Pr_{0.1}Ce_{0.9}O_{2-\delta}/SrTiO_3$ Heterostructures. *ACS Appl. Mater. Interfaces* **11**, 34841–34853 (2019).

18. Lorger, S., Usiskin, R. & Maier, J. Transport and Charge Carrier Chemistry in Lithium Oxide. *J. Electrochem. Soc.* **166**, A2215–A2220 (2019).

19. Srivastava, J. K., Prasad, M. & Wagner, J. B. Electrical Conductivity of Silicon Dioxide Thermally Grown on Silicon. *J. Electrochem. Soc.* **132**, 955–963 (1985).